\documentclass[aps,prd,preprintnumbers,nofootinbib,twocolumn,superscriptaddress]{revtex4-2}
\usepackage[utf8]{inputenc}
\usepackage{amsmath}
\usepackage[english]{babel}
\usepackage{amsfonts}
\usepackage{graphicx}
\usepackage[toc,page]{appendix}
\usepackage{comment}
\usepackage{hyperref}
\usepackage{rotating}
\usepackage{cleveref}
\usepackage{braket}
\usepackage{bbold}
\usepackage{amsmath}
\usepackage{xcolor}
\usepackage{amsthm}
\usepackage[left=2cm,right=2cm,top=2cm,bottom=2cm]{geometry}
\usepackage{amssymb}
\DeclareMathOperator{\tr}{tr}

\newcommand{\I}{\mathrm{i}}
\newcommand{\E}{\mathrm{e}}
\newcommand{\LL}{\mathsf{L}}
\newcommand{\X}{\mathsf{Q}}
\newcommand{\NL}{\mathrm{NL}}
\newcommand{\gf}{\mathrm{gf}}
\newcommand{\Q}{a}
\newcommand{\jj}{j}
\newcommand{\Det}{\mathrm{Det}}
\newcommand{\gh}{\mathrm{gh}}
\newcommand{\MSbar}{$\overline{\text{MS}}\,$}
\newcommand{\gR}{g_{\text{R}}}
\newcommand{\BR}{B_{\text{R}}}
\newcommand{\ZF}{Z_F}
\newcommand{\Nc}{N_{\text{c}}}

\newcommand{\Eqref}[1]{Eq.~\eqref{#1}}

\definecolor{darkpastelgreen}{rgb}{0.01, 0.75, 0.24}
\definecolor{shiraz}{rgb}{0.6,0,0.4}

\newcommand{\dd }{\mathrm{d}}
\newcommand{\Tr }{\mathrm{Tr}}

\newcommand{\Gv}{W^{(2)}_v}
\newcommand{\bfell}{\bar{f}_\ell}

\begin{document}

\title{Background Effective Action with Nonlinear Massive Gauge Fixing}

\author{Holger Gies}
\email{holger.gies@uni-jena.de}
\affiliation{\mbox{\it Theoretisch-Physikalisches Institut, Friedrich-Schiller-Universit{\"a}t Jena,}
	\mbox{\it D-07743 Jena, Germany}}
\affiliation{\mbox{\it Abbe Center of Photonics, Friedrich-Schiller-Universit{\"a}t Jena,}
	\mbox{\it D-07743 Jena, Germany}}
\affiliation{Helmholtz-Institut Jena, Fr\"obelstieg 3, D-07743 Jena, Germany}

\author{Dimitrios Gkiatas}
\email{dimitrios.gkiatas@uni-jena.de}
\affiliation{\mbox{\it Theoretisch-Physikalisches Institut, Friedrich-Schiller-Universit{\"a}t Jena,}
	\mbox{\it D-07743 Jena, Germany}}

\author{Luca Zambelli}
\email{luca.zambelli@bo.infn.it}
\affiliation{\mbox{\it INFN-Sezione di Bologna, via Irnerio 46, I-40126 Bologna, Italy}}

\begin{abstract}

We combine a recent construction of a BRST-invariant, 
nonlinear massive gauge fixing with the background 
field formalism. The resulting generating functional preserves background-field 
invariance as well as BRST invariance of the quantum field manifestly. The 
construction features  BRST-invariant mass parameters for the quantum gauge and 
ghost fields.

The formalism employs a background Nakanishi-Lautrup field which is part of the 
nonlinear gauge-fixing sector and thus should not affect observables. We verify 
this expectation by computing the one-loop effective action and 
the corresponding beta function of the gauge coupling as an example. The 
corresponding Schwinger functional generating connected correlation functions 
acquires additional one-particle reducible terms that vanish on shell. We also 
study off-shell one-loop contributions
 in order to explore the 
consequences of a nonlinear gauge fixing scheme involving a background 
Nakanishi-Lautrup field.

As an application, we show that our formalism straightforwardly 
accommodates nonperturbative information about propagators in the Landau 
gauge in the form of the so-called decoupling solution. Using this 
nonperturbative input, we find evidence for the formation of a
 gluon condensate 
for sufficiently large coupling, whose
	scale is set by the BRST-invariant gluon mass parameter.

\end{abstract}

\maketitle

%\tableofcontents
%\clearpage

\pagenumbering{arabic}

\section{Introduction} \label{s1}

Background fields represent a useful, in some cases essential tool in quantum 
field theory. Background fields can be chosen to represent a specific physical 
environment of interest. In continuum gauge theories, the virtue of background 
fields is that a necessary gauge fixing can be performed such that the 
generating functionals remain manifestly invariant under gauge transformations 
of the background field. For suitably chosen gauges, even the full generating 
functional, e.g., in the form of the effective action for the quantum 
expectation values, inherits this symmetry property from the background field 
\cite{Abbott:1981ke,Abbott:1980hw}. 

Another elegant method to deal with gauge symmetry in continuum 
gauge-fixed computations is based on BRST symmetry, a residual global 
supersymmetry that acts nonlinearly on the gauge and auxiliary degrees of 
freedom \cite{Becchi:1974md,Becchi:1975nq,Tyutin:1975qk}. BRST symmetry  
allows to formulate the constraints for correlation functions imposed by 
gauge invariance in a closed form given by the Zinn-Justin master 
equation~\cite{ZinnJustin:1974mc,ZinnJustin:1975wb}. As a significant 
advantage, the latter can be addressed algebraically with the aid of cohomology 
methods. 

Although a continuum Faddeev-Popov formulation of
	gauge theories, featuring background-gauge and BRST 
	invariance, has proved  extremely useful, 
	especially for scattering amplitude computations,
	it still faces the open challenge of allowing
	for computations of more general physical observable properties
	such as for instance a mass spectrum. These difficulties
	are notoriously related to the phenomena of mass generation and of confinement.	Setting aside nonperturbative issues
	such as the Gribov problem or the Neuberger-zero problem,
	recent studies have argued that effective Lagrangians very similar to those appearing in the Faddeev-Popov method,
	namely the Curci-Ferrari massive deformation thereof,
	are suprisingly accurate as phenomenological tools
	for the description of such phenomena~\cite{Tissier:2010ts,Tissier:2011ey,Reinosa:2013twa,Pelaez:2014mxa,Reinosa:2014ooa,Reinosa:2017qtf,Reinosa:2021}.
	Unfortunately, these effective models do not
	feature a nilpotent BRST symmetry allowing for
	a cohomological construction of unitarity-preserving
	physical Hilbert spaces.
	These studies motivate a renewed interest in
	the open problem of constructing BRST invariant gauge-fixed
	Lagrangians featuring mass parameters for all fields.

In a recent work \cite{Asnafi:2018pre}, 
a class of these BRST-invariant massive models
has been constructed, together with a renormalization group (RG) flow 
equation that offers a route to preserving exact BRST 
symmetry also for nonperturbative approximation schemes. In this way, technical 
challenges posed by other formulations of RG flows based on momentum-space 
regularizations can be overcome 
\cite{Ellwanger:1994iz,Ellwanger:1995qf,Ellwanger:1996wy,Gies:2003dp,
Fischer:2004uk,
Fischer:2008uz,Cyrol:2016tym,Fejos:2017sjl,Bonini:1993sj,DAttanasio:1996tzp,
Litim:1998wk,Litim:1998nf}.

 The main idea of this construction is to 
combine Faddeev-Popov quantization with a nonlinear gauge condition in such a 
way that the regularization procedure (i.e. the mass terms) becomes part of the 
gauge-fixing sector. Additionally, the construction goes along with the 
appearance of an external-field variant of 
the Nakanishi-Lautrup field. In practice, the latter leads to a proliferation of
possible operators that can appear in the BRST-invariant effective action and 
thus higher complexity, see \cite{Asnafi:2018pre} for a computation of several 
gluonic wave-function renormalizations; however, physical observables should 
not be affected by this external field as it is part of the gauge-fixing 
sector.

These observations are a main motivation for the present work in which we 
combine the construction of \cite{Asnafi:2018pre} with the background field 
method in such a way that quantities of physical interest are characterized by 
the full invariance under background transformations whereas the BRST symmetry 
still holds for the quantum gauge fields and its gauge-fixing sector. In this 
way advantages of both methods become available for concrete computations. 

For this, we generalize the nonlinear gauge fixing of \cite{Asnafi:2018pre} 
featuring mass-parameters for the ghost and gluon fields
to the inclusion of a background field preserving all relevant invariances. 
A generalization to nonperturbative 
RG flow equations in the tradition of background field flows 
\cite{Reuter:1993kw,Reuter:1994zn,Reuter:1997gx,Freire:2000mn,Freire:2000bq,
Pawlowski:2005xe,Bridle:2013sra,Becker:2014qya,Dietz:2015owa,Percacci:2016arh,
Safari:2016dwj,Safari:2016gtj,Ohta:2017dsq}, as applied in practice in 
convenient approximation schemes 
\cite{Reuter:1994zn,Reuter:1997gx,Gies:2002af,Braun:2010cy,Eichhorn:2010zc} is 
straightforwardly possible. 
Of course, gauge invariant RG flows do not 
hinge on the background field method. For alternative formulations of 
gauge-invariant flows, see, e.g., 
\cite{Morris:1998kz,Arnone:2001iy,Arnone:2002cs,Branchina:2003ek,
Pawlowski:2003sk,Arnone:2005fb,Morris:2005tv, 
Morris:2006in,Evans:2006eq,Rosten:2010vm, 
Donkin:2012ud,Wetterich:2016ewc,Wetterich:2017aoy,Falls:2020tmj}. Several 
approaches to BRST invariant flows have been constructed and used, e.g., in 
\cite{Igarashi:1999rm,Sonoda:2007dj,Igarashi:2007fw,
Ardalan:2011ri,Lavrov:2012xz,Igarashi:2016gcf,Safari:2016gtj,Barra:2019rhz,
Pawlowski:2020qer,Igarashi:2021zml}; gauge-consistent solutions of 
non-perturbative vertex expansions have recently become accessible in this way 
\cite{Pawlowski:2022oyq}.

As a first application, we perform a nontrivial check by computing the 
flow of the one-loop effective action and the corresponding $\beta$ function of 
the coupling. The result demonstrates explicit 
independence of the external Nakanishi-Lautrup field which serves as a test of 
the required independence of details of the gauge-fixing sector. 
In addition to rediscovering standard results, the BRST-invariant mass 
terms can be used for a controlled investigation of IR phenomena: e.g., the 
perturbative Landau pole of the running coupling can be screened by a 
controlled decoupling of IR modes. The same holds for the unstable 
Nielsen-Olesen mode in constant magnetic fields affecting the effective action. 
Moreover, our formalism straightforwardly accommodates nonperturbative 
information about propagators in the Landau gauge in the form of the so-called 
\textit{decoupling solution} featuring a massive gluon and a massless ghost 
\cite{Sternbeck:2005tk,Aguilar:2007nf,Boucaud:2006if,Cucchieri:2007md,%
Boucaud:2008ky, Dudal:2007cw,Fischer:2008uz, Dudal:2008sp,Bogolubsky:2009dc,% 
Aguilar:2011ux,Sternbeck:2012mf,Aguilar:2015bud,Huber:2018ned,Huber:2020keu,% 
Napetschnig:2021ria}. Using this 
nonperturbative input together with a selfdual background, a simple one-loop 
computation already provides evidence for the formation of a gluon condensate 
for sufficiently large coupling in agreement with results from nonperturbative 
functional RG studies \cite{Eichhorn:2010zc,Horak:2022aqx}. 

The present paper is organized as follows: In Sect. \ref{s1s1} we construct the 
gauge-fixed generating functional of quantum gauge theory analogously to 
\cite{Asnafi:2018pre} but upon the inclusion of a background field and 
preserving background-field invariance. Section \ref{s1s2} is devoted to 
an analysis of a consistency condition on the background Nakanishi-Lautrup 
field. In Sect.~\ref{sec:WvsGamma}, we study the relation between the Schwinger 
functional and the effective action, where the former acquires an additional 
one-particle reducible (1PR) term from our gauge-fixing sector. Section 
\ref{sec:Gamma1L_beta} is focused on the one-loop effective action and contains 
most of our results of phenomenological relevance. Sections 
\ref{s1s4}-\ref{s1s6} investigate the new 1PR term in the Schwinger functional 
in order to analyze potential contributions to scattering amplitudes; we 
develop various approaches to implement the consistency condition arising from 
the gauge-fixing section.
We also discuss the possibility
to treat the background Nakanishi-Lautrup field
as a disorder field,
thus computing quenched averages over it.
We conclude in Sect.~\ref{sec:conc}.

\section{Background Quantization with Fourier Noise} \label{s1s1}
The goal of this section is to quantize the off-shell formulation of pure 
Yang-Mills theory in a BRST- and background-invariant manner. For this, we 
follow Ref.~\cite{Asnafi:2018pre} where the 
contribution of the auxiliary field, which leads to the BRST 
invariant off-shell action, is encoded within a generalized gauge-fixing sector 
via the noise action. Here, we begin by working in $d$-dimensional Minkowski spacetime. However, after performing the gauge fixing procedure we shall Wick rotate the gauge-fixed generating functional to the $d$-dimensional Euclidean spacetime. In addition, we use explicit adjoint indices to label the fields 
and quantities in color space. For instance, the covariant derivative in the 
adjoint representation takes the form
\begin{align}
D_\mu^{ab} = \partial_\mu \delta^{ab} + \bar{g}  f^{acb}A_\mu^c .
\end{align}
Under finite gauge transformations, the gauge field transforms as follows
\begin{align}
A_\mu^\omega =&\ U A_\mu U^{-1} + {\frac{\I}{\bar{g}}} (\partial_\mu U)U^{-1}, \label{e1}\\
U(\omega) =&\ e^{\I\bar{g} \omega^a T^a},
\end{align}
and infinitesimally
\begin{align}
\delta_\omega A_\mu^a =  \partial_\mu \omega^a + g f^{acb} \omega^b A_\mu^c = D_\mu^{ab} \omega^b, \label{e2}
\end{align}
with $\omega(x)$ being the gauge parameter. The gauge 
field lives in the adjoint representation $A_\mu = A_\mu^a T^a $, where 
$(T^a)^{bc} = \I f^{abc}$.
Then, $U(\omega) =   e^{\bar{g}\omega^a f^{acb}}$ corresponds to spacetime dependent  elements of the Lie group $G$.
The components of the field strength tensor $F_{\mu \nu} = F_{\mu \nu}^a T^a$ in the adjoint representation read
\begin{align}
F_{\mu \nu}^a = \partial_\mu A_\nu^a - \partial_\nu A_\mu^a + \bar{g} f^{abc}A_\mu^b A_\nu^c, \label{e3}
\end{align}	
which leads to the pure Yang-Mills action of the form
\begin{align}
S_{\text{YM}} = -\frac{1}{4} F_{\mu \nu}^a F^{a \mu \nu}. \label{eq:Mink_YMact}
\end{align}
Here and in the following, we use a condensed notation: whenever two identical 
color indices both refer to field variables, the summation convention
over these repeated indices is extended to integration over the corresponding 
spacetime points, which are condensed and implicitly associated to
color indices.

For quantization, we start from the vacuum-persistence amplitude in the 
off-shell formulation where we integrate over gauge inequivalent configurations 
using the Faddeev-Popov method with corresponding ghosts $c, \bar{c}$ 
together with a Nakanishi-Lautrup auxiliary field $b$ and a noise field 
$n=n^a 
T^a$. The latter has been introduced in \cite{ZinnJustin:1989mi} and employed 
in the present context in \cite{Asnafi:2018pre}. We write
\begin{align}
Z =  \int{\! \mathcal{D} A \mathcal{D}b \mathcal{D}{c} \mathcal{D} \bar{c} 
\mathcal{D}n\;  e^{\I {\left({S_\text{YM}[A]+ S_\text{gauge} [A, b,c,\bar{c}, n]}\right)}}
}, \label{e5}
\end{align}
where 
\begin{align}
S_{\text{gauge}} [A, c,\bar{c},b] &= S_\text{gf}[A,b] + 
S_{\text{noise}}[b, n] + S_\text{gh}[A, c,\bar{c}] ,\\
S_\text{gf} [A,b] &= - b^a F^a[A] ,\\
S_\text{gh}[A, c,\bar{c}] & =  \bar{c}^a \mathcal{M}^{ab}(A) c^b  .
\end{align} 
Here, $F^a[A]$ denotes the gauge-fixing functional which also determines the 
Faddeev-Popov operator:
\begin{align}
\mathcal{M}^{a b} = \left.\frac{\delta F^a[A^\omega]}{\delta \omega^b}\right|_{\omega=0}  = %\left({\frac{\delta F^a[A^\omega]}{\delta A_\mu^c}}\right) \left({\frac{\delta A_\mu^c}{\delta \omega^b}}\right) =
 \left({\frac{\delta F^a[A]}{\delta A_\mu^c}}\right) D_\mu^{cb}.
 \label{eq:FPop}
\end{align}
Upon integrating out the noise field, the noise action can be translated 
into an action for the Nakanishi-Lautrup field 
\begin{align}
e^{ \I S_{\text{NL}}[b]} = \int \!\mathcal{D}n \; e^{ \I 
S_{\text{noise}}[b,n]} .\label{e6}
\end{align}
Choosing a suitable Gaussian weight for the noise action reproduces the known 
local action which is quadratic in the auxiliary Nakanishi-Lautrup field, 
$S_{\text{NL}}= \frac{\xi}{2} \int b^a b^a$, where $\xi$ denotes the gauge 
paramter, cf.~\cite{Asnafi:2018pre} for details. The corresponding BRST 
invariant action can be translated into its on-shell version by integrating out 
the $b$ field. The bare action of \Eqref{e5}, is 
off-shell BRST invariant
for any $S_{\text{noise}}$,
see the end of this section for more details
on this symmetry.

Whereas Faddeev-Popov quantization fixes the gauge transformations of the 
quantum field $A_\mu$, retaining BRST as a residual global symmetry, the 
background field method can be used in order to maintain the local symmetry of 
background field gauge transformations. For this, we decompose
the gauge field $A_\mu^a$,
\begin{align}
A_\mu^a \to \bar{A}_\mu^a +  \Q_\mu^a ,
\label{e7} 
\end{align}
into a background field $\bar{A}_\mu^a$ and fluctuations which we denote by 
$\Q_\mu^a$. In addition, we also change the gauge-fixing functional in a 
manner that allows to preserve background-field invariance, as shown in \Eqref{e16}. From 
here on, we always use background and fluctuation fields only.
Inserting such a decomposition in the vacuum-persistence amplitude \Eqref{e5},
and coupling external sources $j_\mu$ to the
fluctuations only, we obtain the following generating functional
\begin{align}
Z[j,\bar{A}]=& \int\! \mathcal{D}\Q \mathcal{D}b \mathcal{D}n \; \exp \left\{ \I \left({ S_{\text{YM}}[\Q+\bar{A}] + S_{\text{gf}}[\Q,\bar{A},b]}\right.
\right.\nonumber\\
&\left.{
 \left. {+ S_{\text{noise}}[b,n] + \jj_\mu^a \Q^{a \mu}}\right) }\right\} \; \Delta_{\text{FP}} [\Q, \bar{A}].
 \label{e8}
\end{align}
Here, we have introduced the Faddeev-Popov determinant $\Delta_{\text{FP}} = 
\det \mathcal{M}$ of the corresponding operator $\mathcal{M}^{ab}$ of 
\Eqref{eq:FPop} after integrating out the ghost fields.

Instead of the standard Gaussian weight mentioned above, we choose a Fourier 
weight for the noise action following Ref.~\cite{Asnafi:2018pre},  
\begin{align}
S_{\text{noise}}[b,n] = \left({v^a-b^a}\right) n^a ,\label{e9}
\end{align}
where $v=v^a  T^a$ corresponds to an external scalar
field, which adds to the set of adjoint color fields of the theory. Averaging 
over the noise field, leads to the Nakanishi-Lautrup action of the form
\begin{align}
e^{-S_{\text{NL}}[b]} = \int\! \mathcal{D}n \exp\left\{\I n^a (v^a-b^a)\right\} = \delta{[v^a-b^a]}. \label{e10}
\end{align}
As a consequence, the generating functional using the Fourier weight takes the 
form
\begin{equation}
\begin{aligned}
Z[j,\bar{A}] = \int\!\mathcal{D} \Q &\exp\left\{ \I \left({ S_{\text{YM}}[\Q+\bar{A}] - \, v^a F^a[\Q,\bar{A}]}\right.
\right.\\
&\qquad \quad \left. 
\left.{+ \jj_\mu^a \Q^{a \mu}}\right) \right\} \Delta_{\text{FP}}[\Q,\bar{A}], \label{eq:mink_path}
\end{aligned}
\end{equation}
where a possible additional dependence of the gauge-fixing sector on the 
external field $v$ is assumed but not explicitly indicated.

In order to facilitate the upcoming computations, we perform a Wick rotation of the
gauge-fixed generating functional. Then, the Euclidean path integral takes the 
form
\begin{equation}
\begin{aligned}
Z[j,\bar{A}] = \int\!\mathcal{D} \Q &\exp\left\{ - \left({ S_{\text{YM}}[\Q+\bar{A}] + \, v^a F^a[\Q,\bar{A}]}\right.
\right.\\
&\qquad \quad \left. 
\left.{- \jj_\mu^a \Q_\mu^{a}}\right) \right\} \Delta_{\text{FP}}[\Q,\bar{A}], 
\end{aligned}
\label{e11}
\end{equation}
where the Euclidean Yang-Mills action corresponds to
\begin{align}
S_{\text{YM}} &= \frac{1}{4} F^a_{\mu \nu} F^a_{\mu \nu}. \label{e4}
\end{align}
The Fadeev-Popov determinant is generated by a 
Euclidean functional integral over ghost fields
\begin{align}
\Delta_{\text{FP}}[\Q,\bar{A}] &= \int\!\mathcal{D} c
\mathcal{D} \bar{c} \ \mathrm{e}^{ - S_{\text{gh}}[\Q,\bar{A},c,\bar{c}]},\\
S_{\text{gh}}[\Q,\bar{A},c,\bar{c}]&=-\bar{c}^a \mathcal{M}^{ab}(A) c^b.
\end{align}

We now use the freedom to choose the form of the gauge-fixing condition. In the 
following, it is important to distinguish between two different ways of writing 
the symmetry transformation of the original field $A$ in terms of the 
decomposed fields. In the \textit{quantum gauge transformation}, the background 
field does not transform at all, but the transformation is fully carried by the 
fluctuation field,
\begin{equation}
\begin{aligned}
\bar{A}_\mu^\omega  =&\  \bar{A}_\mu ,\\
\Q_\mu^\omega = U \left({\bar{A}_\mu + \Q_\mu}\right) U^{-1} +& { \frac{\I}{\bar{g}}} 
\left({\partial_\mu U}\right) U^{-1} ,
\end{aligned}
\label{e12}
\end{equation}
or infinitesimally %take the form
\begin{equation}
\begin{aligned}
\delta_\omega^\text{Q} \bar{A}_\mu^a &= 0, \\
\delta_\omega^\text{Q} \Q_\mu^a =  D_\mu^{ab} \omega^b & = \bar{D}_\mu^{ab} 
\omega^b + \bar{g} f^{acb} \omega^b \Q_\mu^c ,
\end{aligned}
\label{e13}
\end{equation}
where ${D}_\mu^{ab}$ is the covariant derivative with respect to the full 
field $A=\bar{A}+{a}$, while $\bar{D}_\mu^{ab}$ is the background 
covariant derivative. 

By contrast, the \textit{background gauge transformation} 
affects both the background and fluctuation field:
\begin{equation}
\begin{aligned}
\bar{A}_\mu^\omega = U {\bar{A}_\mu} U^{-1} +& {\frac{\I}{\bar{g}}} \left({\partial_\mu U}\right) U^{-1}, \\
\Q_\mu^\omega =&\ U \Q_\mu U^{-1}.
\end{aligned}
\label{e14}
\end{equation}
Infinitesimally, we obtain
\begin{equation}
\begin{aligned}
\delta_\omega^\text{B} \bar{A}_\mu^a &=  \partial_\mu \omega^a + \bar{g} f^{acb} \omega^b \bar{A}_\mu^c =  \bar{D}_\mu^{ab} \omega^b,\\
\delta_\omega^\text{B} \Q_\mu^a &= \left[{D_\mu^{ab} - \bar{D}_\mu^{ab}}\right] 
\omega^b =  \bar{g} f^{acb} \omega^b \Q_\mu^c,
\end{aligned}
\label{e15}
\end{equation}
Both transformations add up to the full gauge transformation of the original 
field $A$ in \Eqref{e7}. However, it is the quantum gauge transformation which 
 must be fixed in order to have a well-defined functional 
integral. BRST symmetry will thus correspond to the residual global symmetry of 
the quantum gauge transformation. 

The background field method now consists in choosing a gauge-fixing condition 
for the quantum gauge transformations that is invariant under the background 
gauge transformations. A generic choice are Lorenz-like covariant linear 
gauges, cf.~\cite{Abbott:1980hw,Abbott:1981ke,Dittrich:1985tr}. Here, however, 
we follow \cite{Asnafi:2018pre} and choose a nonlinear gauge-fixing condition 
of the following form:
\begin{align}
F^{a}[\Q,\bar{A}] = \Q_\mu^b \X^{abc}_{\mu \nu} \Q_\nu^c + \LL_\mu^{ab} \Q^b_\mu, \label{e16}
\end{align}
where
\begin{align*}
\X^{abc}_{\mu \nu} &= {\frac{v^a}{2 |v|^2}} \left[{\bar{m}^2 \delta_{\mu \nu}\delta^{bc} - {\frac{1}{\xi}} \bar{D}_\mu^{bd} \bar{D}_\nu^{dc}}\right] ,\\
\LL_\mu^{ab} &= \Big({1 + \frac{\bar{m}_\text{gh}^2}{- \bar{D}^2}}\Big)^{{ac}} \bar{D}_\mu^{cb}.
\end{align*}
The invariance under background transformations is obvious, since the 
differential operators involve background covariant derivatives only, and the 
fluctuation field transforms homogeneously, i.e., color-vector-like, under 
Eqs.~(\ref{e14},\ref{e15}).

The quadratic part in the fluctuations is chosen in such a way that a bare mass 
term for the gauge field is generated by the gauge-fixing action and the 
linear part contributes to the suppression of the IR divergences which arise by 
introducing a ghost mass-like term. Consequently, the gauge-fixing condition, 
\Eqref{e16},
leads to the gauge-fixing action after integration by parts
\begin{align}
S_{\text{gf}}[\Q,\bar{A},v] =\ & {\frac{\bar{m}^2}{2}} \Q_\mu^a \Q_\mu^a + {\frac{1}{2 \xi}}\left(\bar{D}_\mu^{ab} \Q_\mu^b
\right)\!\left(\bar{D}_\nu^{ac} \Q_\nu^c\right) \nonumber\\
&+ v^a \left({1 + \frac{\bar{m}_{\text{gh}}^2}{-\bar{D}^2}}\right) \bar{D}_\mu^{ab}\Q_\mu^b .\label{e17}
\end{align}
The Faddeev-Popov determinant in the nonlinear gauge-fixing condition becomes 
\begin{equation}
\begin{aligned}
&\Delta_{\text{FP}}[\Q,\bar{A}] = \Det
\Bigg\{
\frac{v^a}{2|v|^2} \bigg[2 \bar{m}^2 \Q_\mu^c D_\mu^{cb} 
\\
&- \frac{1}{\xi} \bar{D}_\mu^{cd} \bar{D}_\nu^{de} \Q_\nu^e D_\mu^{cb} - \frac{1}{\xi} \Q_\nu^d \bar{D}_\nu^{de} \bar{D}_\mu^{ec} D_\mu^{cb}\bigg]  \\
&+ \left(1 + \frac{\bar{m}_\text{gh}^2}{-\bar{D}^2}\right) \bar{D}_\mu^{ac} D_\mu^{cb}\Bigg\}, 
\end{aligned}
\label{e18}
\end{equation}
which provides the ghost action
\begin{equation}
\begin{aligned}
&S_\text{gh} [\Q,\bar{A},c,\bar{c}] = -\bar{c}^a \Bigg\{ {\frac{v^a}{2|v|^2}} \bigg[2 \bar{m}^2 \Q_\mu^c 
\\
&- {\frac{1}{\xi}} \bar{D}_\mu^{cd} \bar{D}_\nu^{de} \Q_\nu^e - {\frac{1}{\xi}} \Q_\nu^d \bar{D}_\nu^{de} \bar{D}_\mu^{ec}\bigg] \\
&+ \left(1 + \frac{\bar{m}_\text{gh}^2}{-\bar{D}^2}\right) \bar{D}_\mu^{ac}\Bigg\} \left(D_\mu^{cb}c^b\right).
\end{aligned}
\label{e19}
\end{equation}
The action, apart from the source contribution, is invariant under the 
background gauge 
transformations which transform the remaining fields  
$(\Q, c,\bar{c},v)$ homogeneously. In particular, the 
finite background gauge transformations in addition to \Eqref{e14} 
require:
\begin{align}
%\bar{A}_\mu^\omega &=  U A_\mu U^{-1} - {\frac{\I}{g}} \left({\partial_\mu 
%U}\right) U^{-1} \\
% \Q_\mu^\omega &= U \Q_\mu U^{-1} \\
 v^\omega &=  U v\, U^{-1},\quad
 c^\omega &=  U c\, U^{-1} , \quad
 \bar{c}^\omega &=  U \bar{c}\, U^{-1}. 
\end{align}
The infinitesimal background gauge transformations
correspondingly comprehend
\Eqref{e15} and
\begin{align}
%\delta_\omega^\text{B} \bar{A}_\mu^a &=  \bar{D}_\mu^{ab} \omega^b\\
%\delta_\omega^\text{B} \Q_\mu^a &= - \bar{g} f^{abc} \omega^b \Q_\mu^c \\
\delta_\omega^\text{B} v^a &=  - \bar{g} f^{acb} \omega^b v^c, \,\,
\delta_\omega^\text{B} c^a = - \bar{g} f^{acb} \omega^b c^c,\nonumber\\
\delta_\omega^\text{B} \bar{c}^a &= - \bar{g} f^{acb} \omega^b \bar{c}^c.
\end{align}

Furthermore, the same part of the action, is invariant under the following nilpotent
BRST transformations
\begin{align}
s a_\mu^a &=  {D}_\mu^{ab} c^b, \; \; s \bar{A}_\mu^a = 0 \notag \\
sc^a &= - \frac{\bar{g}}{2} f^{abc} c^b c^c, \; \; s\bar{c}^a = v^a, \; \; s v^a = 0. 
\end{align}
In fact, we can write
\begin{align}
S_\gf[\Q,\bar{A},v]+
S_\gh[\Q,\bar{A},c,\bar{c}]&=s\Psi,\\
\Psi&=\bar{c}^aF^a[a,\bar{A}].
\end{align}
However, the manifest invariance of the action under BRST transformations, for 
nonlinear gauge fixing condition, holds true not only for 
our gauge fixing but also for a wide variety of nonlinear 
gauge fixing conditions~\cite{Das:1982rz,ZinnJustin:1984dt,Giacchini:2019ant}. 
In addition to the BRST transformations we considered, which leave the 
background field invariant, also extended BRST transformations have been 
constructed within the framework of the background field method. In this 
extended version, the 
background field varies by a BRST-closed classical ghost field, initially 
introduced in Ref.~\cite{Grassi:1995wr} and subsequently further 
implemented in the study of different models \cite{Grassi:1999nb, 
Becchi:1999ir, Grassi:1999tp, Ferrari:2000yp}. 
Possible extensions of BRST transformations for	nonlinear gauge-fixing 
conditions are 
	not addressed in this work.

\section{Equations of Motion in the Background Formalism} \label{s1s2}

It is worthwhile to take a closer look at the background field equations of 
motion, as they depend on the external $v$ field entering through the 
nonlinear gauge fixing. For this, we first express the field strength 
tensor which appears in the Yang-Mills action as a function of the background 
field and of the fluctuation according to the 
decomposition of \Eqref{e7},
\begin{align}
F_{\mu \nu}^a &\to \bar{F}_{\mu \nu}^a + \bar{D}_\mu^{ab} \Q_\nu^b - \bar{D}_\nu^{ab} \Q_\mu^b + \bar{g} f^{abc} \Q_\mu^b \Q_\nu^c ,
\label{e20} \\
\bar{F}_{\mu \nu}^a &= \partial_\mu \bar{A}_\nu^a - \partial_\nu \bar{A}_\mu^a + \bar{g} f^{abc} \bar{A}_\mu^b \bar{A}_\nu^c.
\end{align}
Using \Eqref{e20}, the Yang-Mills action, \Eqref{e4}, in the background field 
formalism takes the form
\begin{equation}
\begin{aligned}
&S_{\text{YM}}%[\Q,\bar{A}] 
=  {\frac{1}{4}} \bar{F}_{\mu \nu}^a \bar{F}_{\mu \nu}^a + \bar{F}_{\mu \nu}^a \left({\bar{D}_\mu^{ab} \Q_\nu^b}\right) 
\\
& + {\frac{1}{2}} \left({\bar{D}_\mu^{ab} \Q_\nu^b}\right)\! \left({\bar{D}_\mu^{ac} \Q_\nu^c}\right) - {\frac{1}{2}} \left({\bar{D}_\mu^{ab} \Q_\nu^b}\right)\! \left({\bar{D}_\nu^{ac} \Q_\mu^c}\right) 
\\
& + {\frac{\bar{g}}{2}}  f^{abc}\Q_\mu^b \Q_\nu^c \bar{F}_{\mu \nu}^a + 
\mathcal{O}(a^3). \label{e21}
\end{aligned}
\end{equation}
The second term on the right-hand side  of 
this expression dictates the form
of the background equations of motion 
at vanishing fluctuations. The quadratic parts will be relevant for the 
quantization in the following sections. 

It is convenient to introduce a shorthand notation for
the vector boson's action, i.e.~the total
action at vanishing ghost fields 
\begin{align}
	S_\text{v}[\Q,\bar{A},v]=
	S_\text{YM}[\Q,\bar{A}]+
	S_\text{gf}[\Q,\bar{A},v].
\end{align}
Taking into account the gauge fixing action of \Eqref{e17}, the
vector action reads explicitly
\begin{align}
&S_\text{v}[\Q,\bar{A},v] =  {\frac{1}{4}} \bar{F}_{\mu \nu}^a 
\bar{F}_{\mu \nu}^a \nonumber\\
&+ \left[ \bar{F}_{\mu \nu}^a %\left({\bar{D}_\mu^{ab} \Q_\nu^b}\right) 
+ v^a \left({1 + \frac{\bar{m}_{\text{gh}}^2}{-\bar{D}^2}}\right)
\right] \bar{D}_\mu^{ab}\Q_\mu^b
\nonumber\\
& + \frac{1}{2} \Q_\mu^a \left[
\left(-\left({\bar{D}^{2}}\right)^{ab} + \bar{m}^2\delta^{ab}\right)
\delta_{\mu\nu} +
\left(1-\frac{1}{\xi}\right)\bar{D}_\mu^{ac} \bar{D}_\nu^{cb}
\right] \Q_\nu^b 
\nonumber\\
& + {{\bar{g}}} f^{abc}\Q_\mu^b \Q_\nu^c \bar{F}_{\mu \nu}^a 
+\mathcal{O}(a^3).
\label{e22}
\end{align}
The classical  
background equations of motion are obtained by
\begin{align}
\left.{\frac{\delta S_{\text{v}}[\Q,\bar{A},v]}{\delta 
\Q_\nu^a}}\right|_{\Q \rightarrow 0} = 0, \label{e23}
\end{align}
assuming that the ghost fields vanish in their classical 
configuration.
The resulting classical equations of motion for
the nonlinear gauge fixing 
in the background field formalism can be written as
\begin{align}
\bar{D}_\mu^{ab} \bar{F}_{\mu \nu}^b =
- J_\nu^a[\bar{A},v], \label{e24}
\end{align}
where the deviations from the standard classical equation can be summarized in 
a current
\begin{equation}
	J_\mu^a[\bar{A},v] = \bar{D}_\mu^{ab} \left[{\left({1+ 
\frac{\bar{m}_\text{gh}^2}{-\bar{D}^2}}\right) v}\right]^b.
	\label{eq:defJofv}
\end{equation}
This current is associated to the linear term in the 
gauge-fixing functional. 
However, background-covariant current conservation
\begin{equation}
	\bar{D}_\mu^{ab}J^b_\mu[\bar{A},v]=0, \label{e34a}
\end{equation}
which is implied by \Eqref{e24},
places a restriction on the form of the external field $v^a$, i.e. 
\begin{align}
\left(-\bar{D}^{2} + \bar{m}_\text{gh}^2\right) v^a = 0 .\label{e25}
\end{align}
This last relation must be obeyed by $v^a$  in order for the background equations of motion
\Eqref{e24} to be consistent.

%%%%%%%%%%%%%%%%%%%%%%%%%%%%%%%%%%%%%%%%%%%%%%%%%%%%%%%%%%%%%%%%%%%%%%%%%
\section{Schwinger functional and Effective Action for Nonlinear Gauge Fixing} 
 \label{sec:WvsGamma}
%%%%%%%%%%%%%%%%%%%%%%%%%%%%%%%%%%%%%%%%%%%%%%%%%%%%%%%%%%%%%%%%%%%%%%%%%

The background field formalism provides an elegant path towards constructing 
a gauge-invariant effective action $\Gamma$, being the generating functional of 
one-particle irreducible (1PI) correlation functions. In comparison to standard 
computations in covariant gauges \cite{Dittrich:1985tr}, new structures arise in 
our approach from the nonlinear gauge containing the external $v$ field. In 
order to illustrate these new structures, it is useful to study both the 
effective action as well as the Schwinger functional $W$ which is the 
generating functional for the connected correlation functions. 
Let us start with the 
Schwinger functional defined by
\begin{equation}
	W[\jj,\bar{A}]=\log Z[\jj,\bar{A}],
\end{equation}
where $\jj$ denotes an auxiliary source coupled to the fluctuations.
The effective action is given by the Legendre transform of the 
Schwinger functional, i.e.,
\begin{align}
\Gamma[\Q,\bar{A}] = \sup_{\jj}{\left\{{\jj_\mu^a \Q^a_\mu - 
W[\jj,\bar{A}]}\right\}}. \label{eq:EffAct}
\end{align}
Here, $a_\mu^a$ denotes the so-called classical field conjugate to the
source $j_\mu^a$. By construction, it equals the expectation value of the 
quantum gauge field (in a common abuse of notation also called $a_\mu^a$ 
above) in the presence of the 
source. Corresponding transformations with respect to the ghost fields and a 
ghost source could be introduced but are not necessary for the present 
purpose. From these definitions, we obtain the quantum equation of motion 
\begin{align}
\jj_\mu^a = \frac{\delta \Gamma[\Q,\bar{A}]}{\delta \Q_\mu^a}.
\label{eq:defofj}
\end{align}
Both generating functionals can be represented in terms of a functional 
integral. In the following, we drop the bar of the background field, 
$\bar{A}\to A$, for better readability. Let us start with the Schwinger 
functional,
\begin{align}
\E^{W[\jj,{A}]} = \int\!{\mathcal{D}
	\Q' \; \E^{-S_\text{v}[\Q',{A}] +  \jj_\mu^a \Q_\mu^{\prime a} }\; 
\Delta_\text{FP}[\Q', {A}]}, \label{e28}
\end{align}
where we denoted the integration variables by $\Q'$
to distinguish them from the expectation value $\Q$
introduced above. 
In the following, we concentrate on the one-loop approximation which we obtain 
by expanding the local action to Gaussian order, e.g.,
\begin{eqnarray}
 S_{\text{v}}[\Q',{A}]&\simeq& S_{\text{v}}[0,{A}]+ \frac{\delta  
S_{\text{v}}[0,{A}]}{\delta \Q'_\mu{}^{a}} \Q'_\mu{}^{a}+\frac{1}{2}
\Q'_\mu{}^{a} M^{ab}_{\mu\nu}[{A}]\Q'_\nu{}^{b} \nonumber\\
&& M^{ab}_{\mu\nu}[{A}]= \frac{\delta^2  
S_{\text{v}}[0,{A}]}{\delta \Q'_\mu{}^{a}\delta \Q'_\nu{}^{b}}. 
\label{eq:Svexp}
\end{eqnarray}
Inserting this expansion into \Eqref{e28}, the Gaussian integral can readily 
be performed. We find to 
one-loop order
\begin{eqnarray}
-W_{\text{1L}}[j,A,v]%&=& -W_{\text{1L}}[0,A,v]\nonumber\\
&=&S_\text{v}[A]- 
\ln\Delta_\text{FP}[A] + {\frac{1}{2}} \ln\det M[A] \nonumber \\
&& -W_{\text{source}}[j,A,v],
\label{eq:Schwinger1L}
\end{eqnarray}
where we have introduced
\begin{align}
S_\text{v}[A] & \equiv S_\text{v}[0,A] = \frac{1}{4} F_{\mu \nu}^a F_{\mu 
\nu}^a, \label{eq:bareact}\\
\Delta_{\text{FP}}[A] & \equiv \Delta_{\text{FP}}[0,A]	= 
\det{\!\left[{\left({1+  
 \frac{\bar{m}_\text{gh}^2}{- D^2}}\right)\! \left({D^2}\right)^{ab}}\right]}, 
\label{eq:FPdet}\\
M_{\mu \nu}^{ab}[{A}; \xi] &= \bar{m}^2 \delta_{\mu \nu}^{ab}  - 2\bar{g} f^{abc} F_{\mu \nu}^c - \left({D^2}\right)^{ab} \delta_{\mu \nu} \notag\\
 & \quad +  \left({1 -\frac{1}{\xi}}\right) D_\mu^{ac} D_\nu^{cb}. 
\label{eq:gluonop}
\end{align}
Here $M$ denotes the inverse gluon propagator in the background field. 
The form of $M$ agrees with the standard one for a linear gauge 
condition in the background formalism, 
cf.~\cite{Dittrich:1983ej,Shore:1981mj,Dittrich:1985tr}, except for the gluon mass 
term $\bar{m}^2 \delta_{\mu \nu}^{ab}$, arising from the nonlinear gauge 
fixing. A similar comment applies to the Faddeev-Popov operator. 
Note that the corresponding determinant can be evaluated at vanishing 
fluctuations $\Q'=0$ to one-loop order.
In \Eqref{eq:Schwinger1L}, we have also abbreviated all source-type
contributions to the Schwinger functional 
\begin{align}
	& W_\text{source}[j,A,v] 	\label{eq:Wsource}\\
	&=	\frac{1}{2} (\mathcal{K}_\mu^a[A,v] + j_\mu^a) 
\left({M^{-1}}\right)^{ab}_{\mu \nu}[A] 
	\, (\mathcal{K}_\nu^b[A,v] + j_\nu^b), \nonumber
\end{align}
which is the only part that carries an explicit dependence on the external 
field $v$. This dependence is investigated in the several sections below.
It arises from the background equations of motion, \Eqref{e23} and more 
specifically only from $\frac{\delta S_{\text{gf}}}{\delta {\Q'}_\mu^a}$.
In arriving at \Eqref{eq:Wsource}, we have assumed that the gluonic 
fluctuation 
operator $M$ is invertible, with $M^{-1}$ being its inverse, and introduced the 
abbreviation
\begin{align}
  \mathcal{K}_\mu^a[A,v] \equiv
		-\frac{\delta S_\text{v} [A]}{\delta {\Q'}_\mu^a} = D_\nu^{ab} F_{\nu \mu}^	{b} + 
		J_\mu^a[A,v] . \label{eq:extcurr}
\end{align}
The structure of \Eqref{eq:Wsource} arises from completing the square
in the exponent in the presence of the linear term.

By contrast, the analogous 1PI effective action does not contain such a 
term. 
In order to illustrate the aforementioned claim, we
perform a shift $\Q ' \rightarrow \Q ' + \Q$ in the effective action \Eqref{eq:EffAct} to arrive at 
the integral equation
\begin{equation}
\begin{aligned}
\E^{-\Gamma[\Q,{A}]} = \int\! \mathcal{D}\Q ' 
\; \E^{- S_\text{v}[\Q '+\Q 
,{A}] + \frac{\delta \Gamma[\Q,{A}]}{\delta \Q_\mu^a} \Q'_\mu{}^{a}}
\Delta_\text{FP} [\Q '+\Q ,{A}] .\label{eq:expGamma}
\end{aligned}
\end{equation}
Next, we perform the expansion of the local action similar to 
\Eqref{eq:Svexp} about $\Q$. Again, we assume the presence of the background 
field $A$, but otherwise a vacuum, i.e., a vanishing expectation value $\Q=0$. 
Then, \Eqref{eq:expGamma} takes the form
\begin{align}
\E^{- \Gamma_\text{1L}[0,{A}]} \simeq &\ \int\! \mathcal{D}\Q' \; 
\E^{-S_\text{v}[0,{A}]} \; \E^{-\left({\frac{\delta 
S_\text{v}[0,{A}]}{\delta {\Q'}_\mu^a}  - \frac{\delta 
\Gamma[0,{A}]}{\delta \Q_\mu^a}}\right)  {\Q'}_\mu^{a}}
	\nonumber\\
&\; \quad \quad \times \E^{- \frac{1}{2} {\Q'}_\mu^{a} \left(\!
	\frac{\delta^2 S_\text{v}[0,{A}]}{\delta {\Q'}_\mu^a \delta {\Q'}_\nu^b} 
	\!\right) {\Q'}_\nu^{b}} \; \Delta_\text{FP}[{A}].
\label{eq:Gamma1Lgeneral} 
\end{align}
The effective action can be written as the bare action plus loop 
corrections, $\Gamma=S_{\text{v}}+\Delta \Gamma$. To linear order in the 
exponent of \Eqref{eq:Gamma1Lgeneral}, the contributions from the classical action 
$S_{\text{v}}$ cancel, while the contributions from $\Delta \Gamma$ on the 
right-hand side would induce higher-loop terms and thus can be neglected to 
one-loop order. The result for the one-loop effective action then is
\begin{eqnarray}
\Gamma_{\text{1L}}[A]&=& \Gamma_{\text{1L}}[0,A]\nonumber\\
&=&S_\text{v}[A]- 
\ln\Delta_\text{FP}[A] + {\frac{1}{2}} \ln\det M[A].  \label{eq:Gamma1L}
\end{eqnarray}
We observe that the one-loop effective action acquires a standard form 
consisting of the bare action and ghost and gluon loop contributions in the 
form of functional determinants. The explicit $v$-field dependence has dropped 
out to this order. This illustrates that the power of the background field 
formalism to construct gauge-invariant effective actions is at work also for 
our nonlinear gauge-fixing involving an a priori arbitrary external field 
$v(x)$. 

From a structural perspective, this result can also be understood from the fact 
that the additional term in the Schwinger functional, \Eqref{eq:Schwinger1L}, represents a 
one-particle reducible (1PR) contribution: the two source-like factors 
$\mathcal{K}_\mu^a[A,v]$ are interconnected with a gluon propagator $\sim 
M^{-1}$. By contrast, the effective action generates 1PI correlators by 
construction.
In the background formalism, the difference between the Schwinger functional 
and the effective action on the level of 1PR diagrams can also be constructed 
to higher-loop orders 
\cite{Gies:2016yaa,Karbstein:2019wmj,Karbstein:2021gdi,Ahmadiniaz:2017rrk,
Ahmadiniaz:2019nhk}.

%%%%%%%%%%%%%%%%%%%%%%%%%%%%%%%%%%%%%%%%%%%%%%%%%%%%%%%%%%%%%%
\section{One-loop effective action} %\label{sec:beta}
\label{sec:Gamma1L_beta}
%%%%%%%%%%%%%%%%%%%%%%%%%%%%%%%%%%%%%%%%%%%%%%%%%%%%%%%%%%%%%%

Let us first study the one-loop effective action which gives also direct access 
to the running of the coupling. For obtaining explicit results and in order to 
make contact with the literature, we use the special choice of covariantly 
constant background fields here, defined by 
\begin{equation}
	D_\mu^{ab}F_{\nu\rho}^{b}=0. \label{eq:Covconstcond}
\end{equation}
Further below, we discuss results also beyond this restricted class of fields. 
Covariantly constant fields can be brought into a pseudo-abelian form in which 
the gauge potential can be written as
\begin{equation}
A_\mu^{bc}(x)=\hat{n}^a(T^a)^{bc}\left(-\frac{1}{2}\mathsf{F}_{\mu\nu}
x^\nu\right) ,
\end{equation}
where $\hat{n}^a$ denotes a constant unit vector, $\hat{n}^2=1$, in adjoint 
color space. 

\subsection{Magnetic background and running coupling}
\label{sec:Gamma1L_magnetic}

As a first example, let us choose the constant Abelian field strength 
$\mathsf{F}_{\mu\nu}$ in the form of a constant magnetic field
\begin{align}
\mathsf{F}_{\mu\nu}=B \epsilon^{\perp}_{\mu\nu},
\end{align}
where $\epsilon^{\perp}_{\mu\nu}$ denotes the Levi-Civita symbol in the spatial 
plane orthogonal to the magnetic field direction (in $d=4$ dimensional 
spacetime). For this particular field, the Yang-Mills action reduces to
\begin{equation}
	\frac{1}{4}{F}^a_{\mu\nu}{F}^{a}_{\mu\nu}=\frac{1}{2}B^2.
\end{equation}
For the computation of the color traces, let us introduce the eigenvalues 
$\nu_\ell$ of the the color matrix $(\hat{n}^aT^a)^{bc}$, with 
$\ell=1,\dots,N^2-1$,
and
\begin{equation}
	\bar{B}_\ell=\bar{g}B\nu_\ell.
	\label{eq:diagBl}
\end{equation}
Now, the task is to compute the determinants occurring in \Eqref{eq:Gamma1L} 
which, by virtue of the logarithm, can be rewritten into operator traces. 
Let us start with the gluon contribution. For simplicity, we perform the 
computation in the Feynman gauge $\xi=1$, where the operator $M$ of 
\Eqref{eq:gluonop} simplifies to 
\begin{equation}
M= \bar{m}^2  +	{\mathfrak{D}_\text{\text{T}}}, \quad
\left({\mathfrak{D}_\text{\text{T}}}\right)_{\mu \nu}^{ab}  =  - \delta_{\mu 
\nu} \left({D^2}\right)^{ab} + 2ig F_{\mu \nu}^{ab}, 
	\label{eq:KinT1}
\end{equation}
and ${\mathfrak{D}_\text{\text{T}}}$ denotes the covariant spin-1 Laplacian. 
Writing the one-loop effective action of \Eqref{eq:Gamma1L} as 
$\Gamma_{\text{1L}}= S_\text{v} + \Delta \Gamma_{\text{gh}} +\Delta 
\Gamma_{\text{gl}}$, and subtracting the zero-field limit as an overall 
constant (which can be done for each loop contribution separately), the gluonic 
contribution reads
\begin{eqnarray}
 \Delta \Gamma_{\text{gl}}&=&
 \frac{1}{2}\Tr\left[{\ln\left(
 \frac{\bar{m}^2+\mathfrak{D}_\text{\text{T}}[A]}{\bar{m}^2+\mathfrak{D}_\text{\text{T}}[0]}\right)}\right] \label{eq:DelGamma-gl1}\\
&=&\frac{\Omega_{d}d}{2(4\pi)^{d/2}}
	\int_0^{\infty} \frac{\dd s}{s^{1+d/2}}
  \E^{-\bar{m}^2 s}
	\sum_{\ell=1}^{N^2-1}
	\nonumber\\
	&&\qquad\qquad\times\Bigg[1- 
\frac{s\bar{B}_\ell}{\sinh\left(s\bar{B}_\ell\right)}	- 
\frac{4 s\bar{B}_\ell}{d}\sinh\left(s\bar{B}_\ell\right) \Bigg].\nonumber
\end{eqnarray}
Here, we have kept the dependence on the dimension $d$ and used the propertime 
representation of the logarithm and the 
explicit forms known for the heat kernel of the spin-1 Laplacian (cf.~Appendix 
B of \cite{Gies:2002af}). Equation \eqref{eq:DelGamma-gl1} represents the 
zero-point subtracted but unrenormalized gluon loop. The expression is UV 
divergent corresponding to a log-divergence from the lower bound of the $s$ 
integral.

Correspondingly, the ghost contribution acquires the explicit zero-point 
subtracted form,
\begin{eqnarray}
 \Delta \Gamma_{\text{gh}}&=&-
 \Tr\left[{\ln\left(
\frac{\bar{m}_\text{gh}^2-D^2[A]}{\bar{m}_\text{gh}^2 - 
\partial^2}\right)}\right] \label{eq:DelGamma-gh1}\\
&=&-\frac{\Omega_{d}}{(4\pi)^{d/2}}
	\int_0^{\infty} \frac{\dd s}{s^{1+d/2}}
  \E^{-\bar{m}_\text{gh}^2 s}
	\sum_{\ell=1}^{N^2-1}
	\nonumber\\
	&&\qquad\qquad\times\Bigg[1- 
\frac{s\bar{B}_\ell}{\sinh\left(s\bar{B}_\ell\right)}	 \Bigg],\nonumber
\end{eqnarray}
exhibiting a similar UV log-divergence from the lower bound of the $s$ 
integral. Finally, the bare action reads $S_\text{v}= \frac{1}{2} B^2$ in terms 
of the bare background field $B$. 

Since the effective action is invariant under background transformation by 
construction, covariant geometric objects like the covariant derivative must be 
RG invariant. As the covariant derivative entails the product $\bar{g}A_\mu$ of 
bare quantities, also the product $\bar{g}B$ must be RG invariant and thus can 
equivalently be expressed in terms of the renormalized coupling $\gR$ and the 
renormalized field $\BR$,
\begin{equation}
 \bar{g}B=\gR\BR, \quad \gR=\sqrt{\ZF}\bar{g}, \quad \BR= \frac{B}{\sqrt{\ZF}},
\label{eq:renquant}
 \end{equation}
where $\ZF$ denotes the wave function renormalization of the background field. 
In terms of the renormalized field, we can now write the one-loop action as
\begin{eqnarray}
\Gamma_{\text{1L}}&=& \frac{\Omega_{d}}{2} \BR^2 + \Delta \Gamma_{\text{gl}} + 
\Delta \Gamma_{\text{gh}} - \frac{\Omega_{d}}{2} (1-\ZF) \BR^2\\
&=& \frac{\Omega_{d}}{2} \BR^2   + \Delta \Gamma_{\text{gl,R}} + 
\Delta \Gamma_{\text{gh,R}},
\label{eq:Gamma1Lren}
\end{eqnarray}
where we have chosen $\ZF$ such that the log-divergences in the bare loop 
contributions $\Delta \Gamma_{\text{gl/gh}}$ are canceled, and 
finite corrections $\Delta \Gamma_{\text{gl/gh,R}}$ are left.
 More specifically, in the present scheme, we choose
\begin{align}
\ZF=1+ \frac{\Nc}{6(4\pi)^{d/2}} \gR^2 \int_0^\infty \frac{\dd s}{s^{-1+d/2}} \notag
\left[{ (24-d) e^{-\bar{m}^2s}}\right. \\
\left.{ + 2 e^{-\bar{m}_{\text{gh}}^2 s} }\right],
\label{eq:ZF}
\end{align}
where we have made use of the fact that $\sum_\ell \nu_\ell^2=C_2(G)=\Nc$ 
corresponds to the 2nd Casimir of the gauge group $G$, being the number of 
colors for $G=\mathrm{SU}(\Nc)$. The wave function renormalization carries the UV 
divergence which can be regularized in different ways. It is instructive to study 
two examples more explicitly. For simplicity -- and independently of the scheme 
-- we use a common scale $k$ for the gluon and the ghost masses in the 
following,
\begin{equation}
 k^2= \bar{m}^2= \bar{m}_{\text{gh}}^2.
 \label{eq:k}
\end{equation}
Generalization to arbitrary and independent choices are straightforward. 
Let us start by making contact with the \MSbar\ scheme. For this, we use the 
analytic continuation in the variable $d$ of the propertime integral in 
\Eqref{eq:ZF}, and introduce the dimensionless 
renormalized coupling
\begin{equation}
 g^2= \gR^2\, \mu^{d-4}
 \label{eq:gq}
\end{equation}
with $\mu$ being an arbitrary (renormalization) scale. Choosing 
$d=4-2\epsilon$ and expanding in $\epsilon$, yields
\begin{equation}
 \ZF=1- \frac{22\Nc}{6(4\pi)^2} g^2 \left(\frac{-1}{\epsilon} + \ln 
\frac{k^2}{\mu^2} -\ln 4\pi + \gamma - \frac{1}{11}\right),
\label{eq:ZFdimreg}
\end{equation}
ignoring higher orders in $\epsilon$. From here, we can immediately deduce the 
one-loop beta function that governs the scale dependence of the renormalized 
coupling,
\begin{equation}
 \beta_{g^2}= - \frac{\partial}{\partial \ln \mu} g^2 = - \frac{22\Nc}{3} 
\frac{g^4}{(4\pi)^2}.
\label{eq:betaMSbar}
\end{equation}
From \eqref{eq:ZFdimreg}, it is obvious that we could equivalently study the 
dependence of the renormalized coupling on the gluon and ghost mass scale $k$, 
serving as an IR regulator, in order to explore the behavior of the theory 
at different scales. I.e., we obtain the same beta function from $\beta_{g^2}= 
k \frac{\partial}{\partial k} g^2$. 

As an alternative to dimensional regularization and the \MSbar\ scheme, we can 
regularize the $s$ integral in \Eqref{eq:ZF} by a propertime regulator at the 
lower bound directly in $d=4$ dimensions. The corresponding integral then 
yields
\begin{equation}
 \int_\frac{1}{\Lambda^2}^\infty \frac{\dd s}{s} \, e^{-k^2s} = 
\Gamma\left({0,\frac{k^2}{\Lambda^2}}\right)= -\ln \frac{k^2}{\Lambda^2} -\gamma + 
\mathcal{O}(k^2/\Lambda^2),
\label{eq:PTreg}
\end{equation}
where $\Lambda$ denotes a UV momentum cutoff scale. In the limit 
$\Lambda\to\infty$, we reobtain \Eqref{eq:betaMSbar} from the construction 
$\beta_{g^2}= 
k \frac{\partial}{\partial k} g^2$. However, we can also keep $\Lambda$ finite, 
defining an explicitly mass-dependent regularization scheme. In this case, we 
obtain for the beta function
\begin{equation}
 \beta_{g^2}=  k \frac{\partial}{\partial k} g^2 = - \frac{22\Nc}{3} 
\frac{g^4}{(4\pi)^2} e^{-\frac{k^2}{\Lambda^2}}.
\label{eq:betaPTmass}
\end{equation}
Whereas the standard result is rediscovered for $\Lambda\to\infty$ (or 
$k\to0$), the beta function approaches zero if the mass scale $k$ surpasses the 
UV scale. This is a typical threshold behavior characterizing the decoupling of 
modes.

\begin{figure}[t]
\centering
\includegraphics[width=0.48\textwidth]{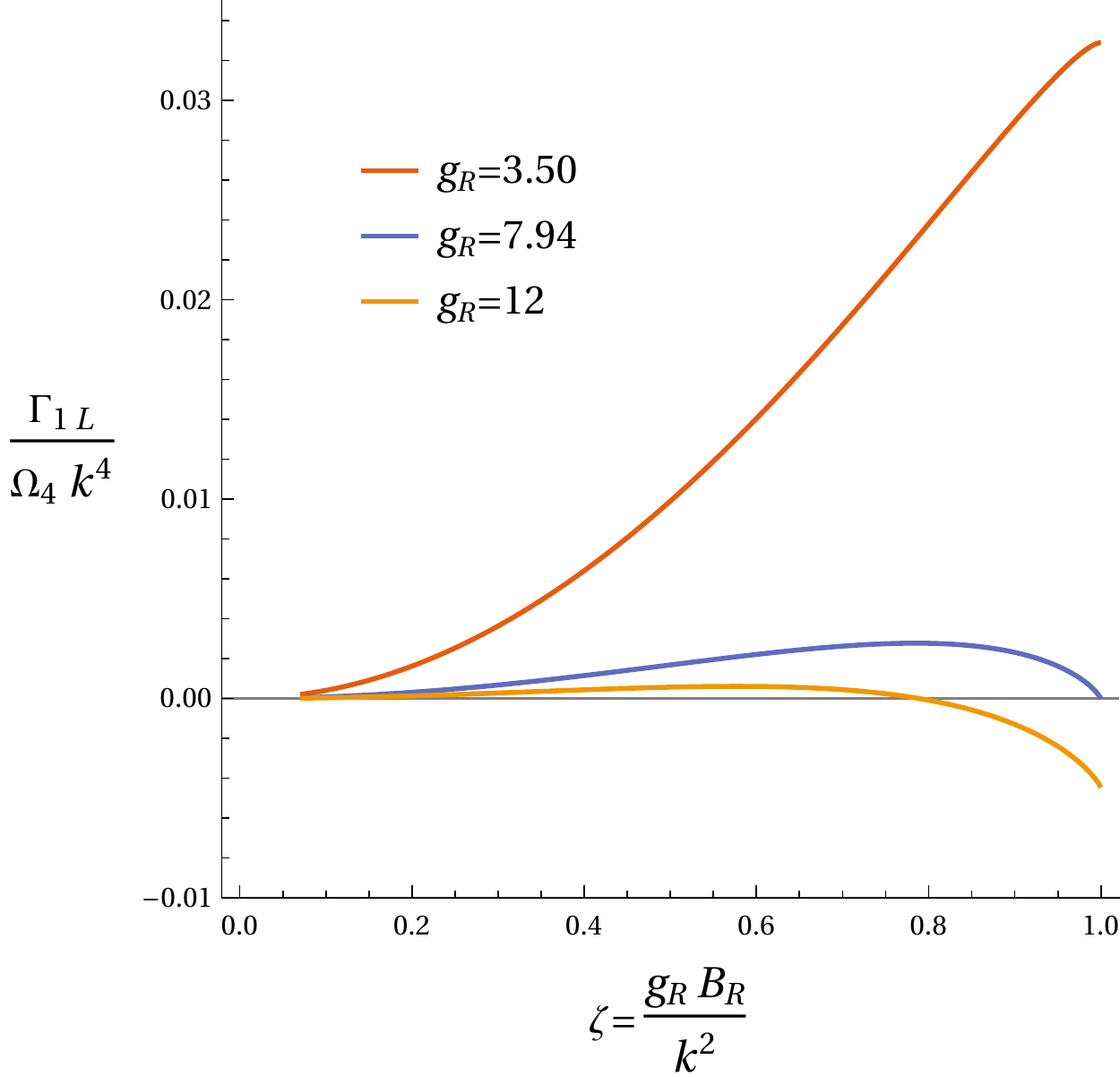}
\caption{Numerical result for the dimensionless action density, as determined in 
\Eqref{eq:Gamma1Lrenorm}, in terms of the dimensionless parameter $\zeta = 
\frac{g_\text{R} B_\text{R}}{k^2}\leqslant 1$, for various values of the renormalized 
coupling constant. Increasing the renormalized coupling $g_\text{R}$ results in 
an increasing but finite contribution of the quantum part of the one-loop 
effective action eventually dominating the classical one. }  
\label{fig:dimlessLagr}
\end{figure}

Let us now come back to the renormalized loop contributions to the one-loop 
effective action. After subtraction of the counterterms, the finite result for 
the action density reads in $d=4$ dimensions (using for simplicity $B_\ell= 
\gR \BR \nu_\ell$ and $k^2=\bar{m}^2=\bar{m}_{\text{gh}}^2$),
\begin{align}
 \frac{\Gamma_{\text{1L}}}{\Omega_4}&= \frac{1}{2} \BR^2 
\label{eq:Gamma1Lrenorm}\\
 &+\frac{1}{(4\pi)^{2}}
	\int_0^{\infty} \frac{\dd s}{s^{3}}
  \E^{-k^2 s}
	\sum_{\ell=1}^{N^2-1}
	\nonumber\\
	&\times\Bigg[1+ \frac{11(sB_\ell)^2}{6} - 
\frac{s{B}_\ell}{\sinh\left(s{B}_\ell\right)}	- 
{2s{B}_\ell}\sinh\left(s{B}_\ell\right) \Bigg].\nonumber
\end{align}
Incidentally, this gluonic contribution is finite only for 
$k^2\geq {B}_\ell$. Otherwise, it diverges due to the last term. The 
physical origin is an unstable 
(Nielsen-Olesen) mode developed by the spin-1 Laplacian in a constant magnetic 
field owing to the paramagnetic spin-field coupling 
\cite{Nielsen:1978rm,Nielsen:1978nk}. 
Attempts to resolve this instability have lead to QCD vacuum models of magnetic 
flux tubes of finite spatial extent (``Spaghetti vacuum'') 
\cite{Nielsen:1978rm,Nielsen:1978nk,Nielsen:1979xu,Ambjorn:1979xi,%
Ambjorn:1980ms,Chernodub:2012mu}.   
On a more formal level, this has been dealt with in expressions like 
\Eqref{eq:Gamma1Lrenorm}, e.g., by analytic continuations of the $s$ integral 
contour over the last term, yielding imaginary contributions 
\cite{Schanbacher:1980vq,Gies:2000dw}. These can be interpreted as decay 
probabilities of the 
constant magnetic field towards an inhomogeneous ground state.  
In our formulation, this instability is cured in the presence of a sufficiently 
large BRST invariant mass term $\sim k$ for the gluonic modes. This mass term 
simply screens the unstable mode, provided that $k^2\geq {B}_\ell$. In 
the 
corresponding range, we plot the resulting dimensionless action density for 
various values of the renormalized coupling in terms of the dimensionless 
parameter $\zeta=\frac{g_\text{R} B_\text{R}}{k^2}$  
for the gauge group SU(2) 
($\nu_\ell=-1,0,1$) in Fig. \ref{fig:dimlessLagr}.
For small couplings, the classical contribution is dominating and quadratically 
increasing  with 
the field strength. Modifications become visible only near the 
validity boundary $\zeta = \frac{g_\text{R} B_\text{R}}{k^2}= 1$, 
where the unstable 
mode contributes strongest, cf.~upper line for $\gR=3.5$, i.e. 
$\frac{\alpha_\text{R}}{4\pi}\simeq 0.08$. Extending the one-loop result ad hoc to even 
larger values of the coupling, the one-loop contribution becomes of similar 
order as the classical one near $\gR=7.94$, i.e. 
$\frac{\alpha_\text{R}}{4\pi}\simeq 0.4$, 
cf.~middle line in Fig.~\ref{fig:dimlessLagr}. 
For even larger values of the coupling, the classical trivial ground state no 
longer remains the global minimum as is visible for the lowest line 
for $\gR=12$, i.e. $\frac{\alpha_\text{R}}{4\pi}\simeq 0.9$. The low-lying modes in the 
gluonic spectrum drive the system towards a state of nonvanishing 
field expectation 
value which may be taken as an indication for gluon condensation in the present 
setting.

Note that this follows from extrapolating the one-loop contribution to large 
coupling within the full validity domain of the one-loop effective action as 
our gauge-fixing allows to rigorously control the unstable-mode contribution. 
This is different from the conventional one-loop reasoning, where an analytic 
continuation of the gluon determinant is needed in order to arrive at a 
well-defined result. In the latter case,  the final result is dominated by the 
one-loop counter-term $\sim \frac{11}{6} s^2 B^2$ inside the propertime 
integral, and the one-loop action acquires the conventional $B^2 \ln 
\frac{B^2}{k^2}$ form, which also exhibits a nontrivial minimum 
\cite{Savvidy:1977as}.

\subsection{Selfdual background and gluon condensation}
\label{sec:Gamma1L_selfdual}

Addressing the generation of a gluon condensate requires a fully 
non-perturbative method. For instance, clear evidence is provided by an RG flow 
computation based on non-perturbative propagators \cite{Eichhorn:2010zc}, 
yielding satisfactory agreement with phenomenological estimates from spectral 
sum rules. In the present perturbative setting, it is nevertheless 
straightforwardly possible to find further indications for the onset of gluon 
condensation. For this, we use a Euclidean selfdual background field, e.g., 
with the choice
\begin{equation}
 \mathsf{F}_{12}=\mathsf{F}_{34}=f=\text{const.}, \label{eq:defSD}
\end{equation}
for the abelian field strength. Using conventions analogous to the preceding 
subsection, the classical Yang-Mills action reduces to 
$\frac{1}{4}F_{\mu\nu}^aF_{\mu\nu}^a=f^2$, and the one-loop effective action 
can be expressed in terms of the variable
\begin{equation}
 \bfell= \bar{g}f \nu_\ell. \label{eq:fell}
\end{equation}
The advantage of the selfdual background is that the corresponding spectrum of 
the gluonic transversal fluctuation operator $\mathfrak{D}_{\text{T}}$ does not 
have an unstable mode, but features a double zero mode \cite{Leutwyler:1980ma}. 
Though this removes the instability encountered in the preceding subsection, 
the treatment of the zero mode still requires some care. 

As a second ingredient, we choose the ghost mass term to vanish 
$\bar{m}_{\text{gh}}=0$ while we keep a finite gluon mass $\bar{m}>0$. In fact, 
this mimicks the so-called \textit{decoupling solution} known from the 
nonperturbative study of gluon and ghost propagators 
\cite{Sternbeck:2005tk,Aguilar:2007nf,Boucaud:2006if,Cucchieri:2007md,%
Boucaud:2008ky, Dudal:2007cw,Fischer:2008uz, Dudal:2008sp,Bogolubsky:2009dc,% 
Aguilar:2011ux,Sternbeck:2012mf,Aguilar:2015bud,Huber:2018ned,Huber:2020keu,% 
Napetschnig:2021ria}. A parametrization of these nonperturbative 
propagators in terms of massive gluons but massless ghosts works surprisingly 
successfully in phenomenological applications 
\cite{Tissier:2010ts,Tissier:2011ey,Reinosa:2013twa,Pelaez:2014mxa, 
Reinosa:2014ooa,Reinosa:2017qtf,Reinosa:2021}.  

Analogously to the computation for the magnetic case in the preceding 
subsection, we can compute the corresponding determinants for the selfdual case 
using the explicit forms for the heat kernel (cf. \cite{Eichhorn:2010zc}). 
Equivalently, the one-loop renormalization can be performed, yielding the same 
result for the running coupling. Here, we concentrate on the final form of the 
one-loop effective action in terms of the analogously renormalized field 
strength parameter $f_\text{R}$. Using the abbreviation $f_\ell= g_\text{R} 
f_{\text{R}} \nu_\ell$, the resulting renormalized one-loop action density 
reads
\begin{align}
 \frac{\Gamma_{\text{1L}}}{\Omega_4}&= f_\text{R}^2 
\label{eq:Gamma1LrenormSD}\\
 &+\frac{1}{(4\pi)^{2}}		
  \int_0^{\infty} \frac{\dd s}{s^{3}} \sum_{\ell=1}^{N^2-1}
 	\nonumber\\
	&\times\Bigg\{ \E^{-\bar{m}^2 s}
\Bigg[2+ \frac{11(sf_\ell)^2}{3} - 
\frac{2(s{f}_\ell)^2}{\sinh^2\left(s{f}_\ell\right)}	- 
{4(s{f}_\ell)^2}\Bigg]\nonumber\\
&\qquad - \left(1- \frac{(f_\ell s)^2}{\sinh^2(f_\ell s)} 
\right)\Bigg\}.\nonumber
\end{align}
Here, we have subtracted the counterterms completely within the gluon loop 
terms corresponding to the terms in square brackets; a naive separate 
subtraction of the gluon and ghost loops would artificially induce an IR 
divergence in the ghost term, and also render the wave function renormalization 
IR divergent. By contrast, the present subtraction prescription is a pure UV 
subtraction and renders the one-loop action UV and IR finite. 

\begin{figure}[t]
\centering
\includegraphics[width=0.48\textwidth]{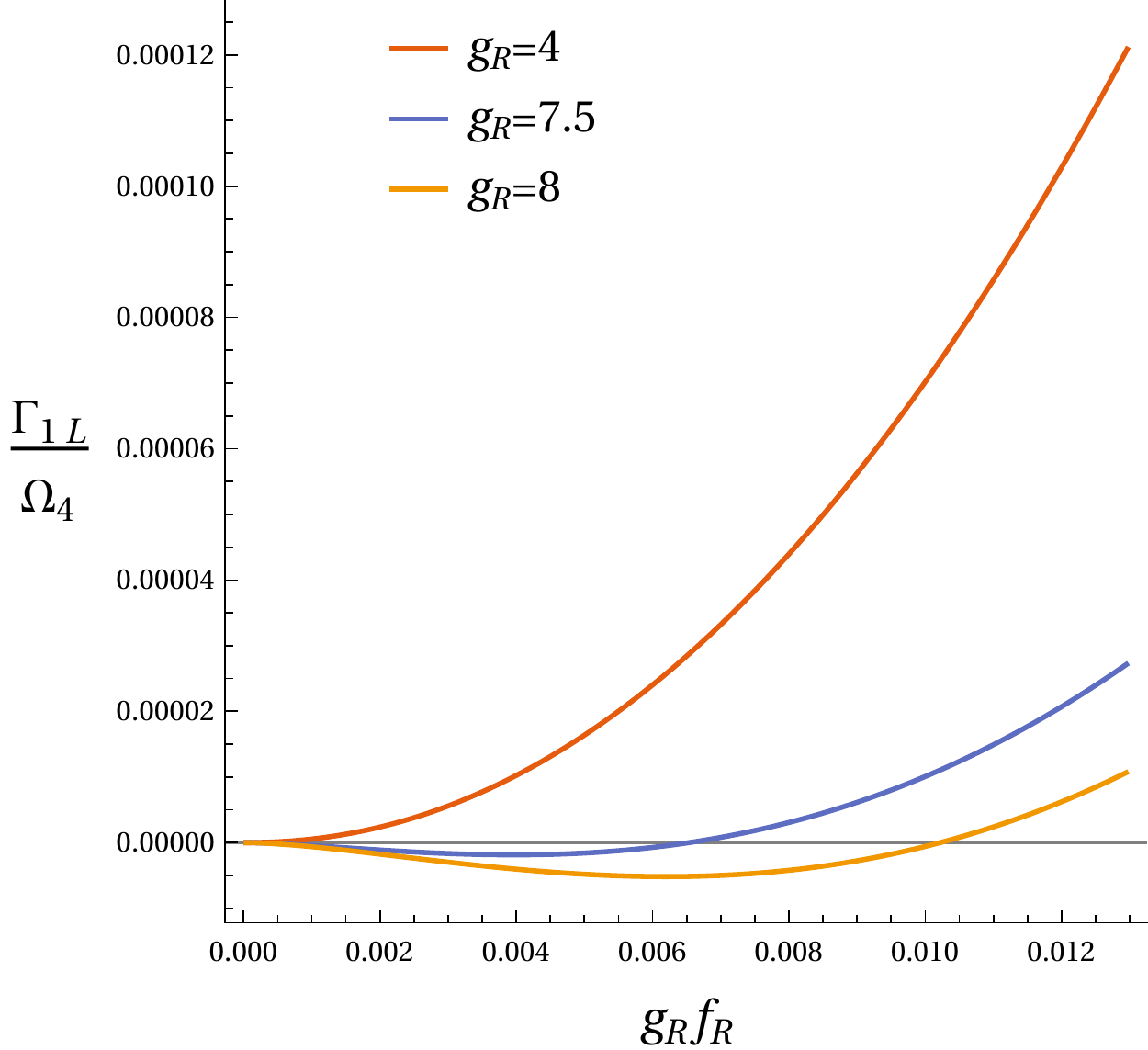}
\caption{Action density for a selfdual background field,  
\Eqref{eq:Gamma1LrenormSD}, in terms of the  parameter $
g_\text{R} B_\text{R}$.
All quantities are plotted in units of the gluon mass which we choose as 
$\bar{m}^2 = 1$ here.
An increase of the renormalized coupling beyond a critical coupling 
$g_{\text{cr}}\simeq 5$ results in the appearance of a nontrivial minimum which 
can be taken as an indication for a gluon condensate. }  
\label{fig:dualbackgroundcondensate}
\end{figure}  

\begin{figure}[t]
\centering
\includegraphics[width=0.48\textwidth]{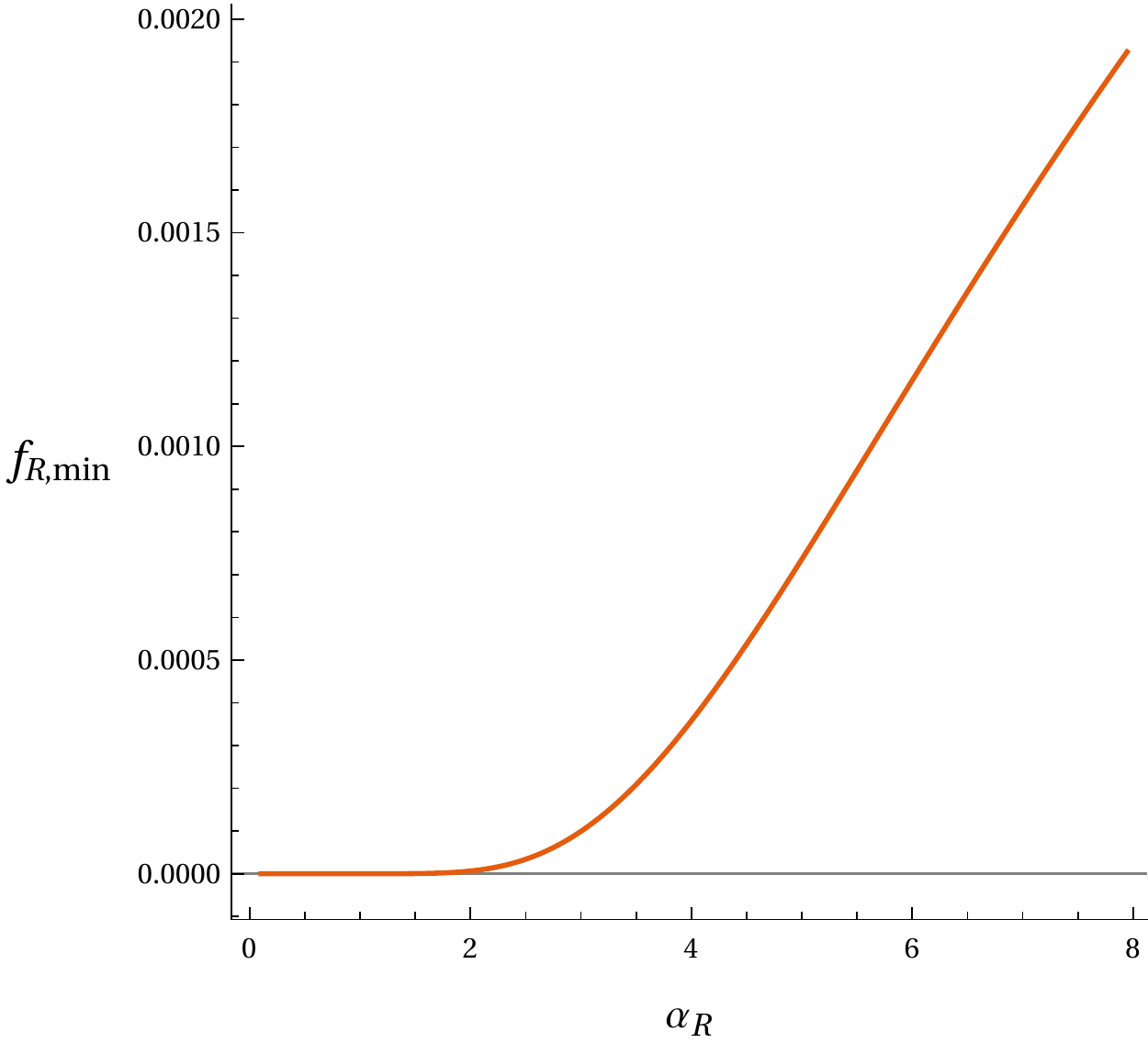}
\caption{Gluon condensate $f_{\text{R,min}}$, in units of the gluon mass, corresponding 
to the nontrivial minimum of the action density as a function of the 
 renormalized coupling $\alpha_\text{R} = \frac{g_\text{R}^2}{4 \pi}$.
Our result shows a continuous increase of the condensate beyond a 
critical coupling $\alpha_\text{R} > 
\alpha_\text{cr}\simeq 2$. The transition to a condensate phase is 
continuous.}   
\label{fig:strongcouplingPlot}
\end{figure} 

The last term of the decomposition of the gluon loop in square brackets $\sim 
4(sf_\ell)^2$ corresponds to the zero-mode contribution. The latter introduces 
a subtlety: as is well known, there exists a nontrivial IR-UV interplay in the 
presence of zero modes \cite{Dunne:2002ta}. Though the zero mode is clearly an 
IR feature of the spectrum, it can affect the strong-field limit of effective 
actions which are generically dominated by the UV properties of the theory. In 
the present case, the zero mode indeed spoils the large-field asymptotics, in 
the sense that this last term in square brackets leads to a  
large-field asymptotics $\sim - f^2 \ln f^2/\bar{m}^4$ unbounded from below. 

Since the zero mode does not contribute to the small-field behavior (it is 
subtracted by the counterterm), this artifact can be cured easily: we may 
multiply the zero-mode contribution with an IR-regularizing function that 
suppresses its large-field asymptotics, for instance, by the replacement 
$-4(sf_\ell)^2 \to -4(sf_\ell)^2 \exp(-s^2/L^4)$. 
Here, $L\gg 1/\bar{m}, \; 
1/\sqrt{f}$ can be thought of as a generic IR length scale over which the 
homogeneous selfduality assumption persists. 

For the following study, the details of this IR regularization are not 
relevant. It suffices to know that a suitable cure of this artifact exists. 
Here, we concentrate on the properties of the effective action in the 
small-field region which is neither affected by the zero mode nor by the 
details of its regularization.  

In Fig.~\ref{fig:dualbackgroundcondensate}, we depict the 
resulting one-loop effective action density for the case of SU(2) as a function 
of the selfdual field strength in units of the gluon mass parameter for 
increasing values of the coupling.  We observe that the   
corresponding action density
is dominated by the classical part for small renormalized couplings, cf.~the line 
for $g_\text{R}=4$, i.e.~$ \alpha_\text{R}=\frac{g_{\text{R}}^2}{4\pi} \simeq 
1.3$ which persists up 
to the critical coupling $g_{\text{cr}}\simeq 5$ (or 
$\alpha_{\text{cr}} \simeq 2$). Further increase of the 
renormalized coupling results in the development of a nontrivial vacuum 
expectation value, cf. lines for $g_\text{R} = 7.5$  ($\alpha_\text{R} 
\simeq 4.5 $) and $g_\text{R} = 8$ ($\alpha_\text{R} \simeq 
5.1 $). The transition to the condensate phase at the 
critical coupling $\alpha_{\text{cr}}\simeq 2$ is continuous and the subsequent 
increase proceeds almost linearly with $\alpha_{\text{R}}$, see 
Fig.~\ref{fig:strongcouplingPlot}.

On the one hand, our approach appears capable of qualitatively addressing 
the question of gluon condensation using some nonperturbative input based on 
the decoupling solution for fully dressed propagators. On the other hand, our 
quantitative result for the value of the condensate depends on the choice for 
the gluon mass parameter $\bar{m}$ and further input for the IR behavior of the 
coupling. Choosing as an example $\bar{m}\simeq 1$GeV as a typical hadronic 
scale and the coupling in the range $\alpha_\text{R}\simeq 2 \dots 8$ does, 
however, not lead to a value of the condensate that would match with 
phenomenological estimates \cite{Narison:2009vy}; the latter would correspond to 
$f_{\text{min}}\bar{m}^2 \simeq 0.21 \text{GeV}^2$ in our conventions.
This poor quantitative accuracy is most likely
due to the one-loop approximation adopted here.

%%%%%%%%%%%%%%%%%%%%%%%%%%%%%%%%%%%%%%%%%%%%%%%%%%%%%%%%%%%%%%%%%%%%%%%%%%
\section{Schwinger functional source term for covariantly constant 
backgrounds} \label{s1s4}
%%%%%%%%%%%%%%%%%%%%%%%%%%%%%%%%%%%%%%%%%%%%%%%%%%%%%%%%%%%%%%%%%%%%%%%%%%

To one-loop order, our nonlinear gauge fixing induces a structural difference 
to conventional formulations in the form of the 1PR contribution to the 
Schwinger functional, cf.~\Eqref{eq:Wsource}. As the Schwinger functional 
generates all connected diagrams, contributing to $S$-matrix elements, this 
additional term is of general interest and is explored from different 
perspectives in this and the following sections. We confine ourselves to 
vanishing external sources $j=0$,
see \Eqref{eq:defofj}, such that \Eqref{eq:Wsource} acquires the form
\begin{equation}
	 W_\text{source}[A,v] 	
	=	\frac{1}{2} \mathcal{K}_\mu^a[A,v] 
\left({M^{-1}}\right)^{ab}_{\mu \nu}[A] 
	\, \mathcal{K}_\nu^b[A,v] , \label{eq:WsAV}
\end{equation}
where $\mathcal{K}$ has been defined in \eqref{eq:extcurr}. A priori, 
$W_{\text{source}}$ depends independently on the background field $A$ as well 
as on the external $v$ field, even though these two fields are connected by a 
constraint, cf.~\Eqref{e25}. Whereas the background field has a natural 
physical interpretation, the $v$ field is part of the gauge fixing, hence, we 
expect that it does not contribute to observables.

At this point, we still have the freedom to choose a condition for the 
form of 
the background $A$ and external field $v$. In the present section, we focus on 
covariantly constant fields, satisfying \Eqref{eq:Covconstcond}. Provided the 
classical equations of motion are fulfilled, this directly implies that the 
current $J[A,v]$ carrying the $v$ field dependence, cf.~\Eqref{eq:defJofv}, 
also has to vanish according to \Eqref{e24}. The immediate conclusion is that 
the source term \Eqref{eq:WsAV} vanishes identically (assuming the absence of 
relevant zero modes of $M$). This proves that -- under these assumptions -- the 
one-loop $S$-matrix elements are independent of the $v$ field of the 
gauge-fixing sector as expected.

However, it is instructive to relax these assumptions and keep the 
discussion slightly more general. From a physical perspective, the potential 
presence of unstable modes for covariantly constant fields, such as the 
pseudoabelian magnetic case used above, suggests to consider the 
covariantly-constant-field assumption as a local approximation to fields varying 
sufficiently slowly in space and time. We still assume vacuum conditions, i.e., 
the absence of external currents $\jj=0$, such that an on-shell stable 
background field configuration should be covariantly constant. 
However we no longer assume
that the classical equations of motion \Eqref{e24} are
fulfilled. 
While we now take 
$J[A,v]$ as potentially non-vanishing, consistency of the equation of motion 
requires $J$ to be covariantly conserved, $D_{\mu}J_{\mu}=0$,
which we keep as an assumption 
also when $A$ is off-shell.

In addition, in this section
we consider a more
general form of the souce
which is only assumed to be 
covariantly longitudinal,
\begin{align}
J_\mu^a[A,v]= D_\mu^{ab}[A] \chi^b, \label{eq:longitudinalcurrent}
\end{align} 
where $\chi^a$ is an arbitrary auxiliary quantity.
The special form of the source
in \Eqref{eq:defJofv} which is specific
for a Fourier-noise implementation of
the background-covariant gauge fixing 
can be recovered by appropriately choosing $\chi$.

For these choices, the source contribution 
\Eqref{eq:Wsource}
becomes
\begin{align}
W_{\text{source}}[A,v] =&\ \frac{1}{2} J_\mu^a[A,v] 
\left({M^{-1}}\right)^{ab}_{\mu \nu}[A] J_\nu^b[A,v].
  \label{e41}
\end{align}
For the concrete analysis, let us consider the gluonic mass parameter 
$\bar{m}$ to be sufficiently large. Also, we keep the gauge-fixing parameter 
$\xi$ general. In view of \Eqref{eq:WsAV} we need the inverse gluonic 
fluctuation operator. For this, let us rewrite \Eqref{eq:gluonop} as 
\begin{align}
M_{\mu \nu}^{ab} [A] = \bar{m}^2 \delta_{\mu \nu}^{ab} + \mathcal{Q}_{\mu 
\nu}^{ab} [A] 
%= \bar{m}^2 \left({\delta_{\mu \nu}^{ab} + \frac{1}{\bar{m}^2} 
%\mathcal{Q}_{\mu \nu}^{ab}[A]}\right) 
,\label{e39}
\end{align}
where 
\begin{equation}
\mathcal{Q}_{\mu \nu}^{ab}[A] = 2\bar{g} f^{abc} F_{\mu \nu}^c - 
\left({D^2}\right)^{ab} \delta_{\mu \nu} +  \left({1 -\frac{1}{\xi}}\right) 
D_\mu^{ac} D_\nu^{cb}.
\end{equation}
The gluonic fluctuation operator and its inverse, schematically are given by
\begin{align}
M & = \bar{m}^2 \left[{\mathbb{1}+ \bar{m}^{-2} \mathcal{Q}}\right],\quad 
%\notag\\
M^{-1} = \bar{m}^{-2} \left[{\frac{1}{1+\bar{m}^{-2} \mathcal{Q}}}\right].  
\label{eq:LGME}
\end{align}
As we assume the gluonic mass to be sufficiently large, we expand the gluon 
propagator in powers of the operator $\mathcal{Q}$,
\begin{align}
\left({M^{-1}}\right)^{ab}_{\mu \nu} = - \sum_{n=0}^{\infty} 
{\left[{\left({\frac{i}{\bar{m}}}\right)^{2n+2} 
\left({\mathcal{Q}^n}\right)_{\mu \nu}^{ab}}\right]}, \label{e40}
\end{align}
where  $ 
\left({\mathcal{Q}^n}\right)_{\mu \nu}^{ab} = \underbrace{\mathcal{Q}^{ac}_{\mu 
\rho} \mathcal{Q}^{cd}_{\rho \sigma} \cdots \mathcal{Q}_{\kappa \tau}^{er} 
\mathcal{Q}^{rb}_{\tau \nu}}_{\text{$n$ times}}$ was used. 
It is useful to decompose the operator $\mathcal{Q}_{\mu \nu}$ in terms of the 
longitudinal $\mathfrak{D}_\text{L}$ and transversal $\mathfrak{D}_\text{T}$  
kinetic operators
\begin{align}
	\mathcal{Q}_{\mu \nu}^{ab} [A] &= 
\left({\mathfrak{D}_\text{\text{T}}}\right)_{\mu \nu}^{ab} - 
\left({1-\frac{1}{\xi}}\right) \left({\mathfrak{D}_\text{\text{L}}}\right)_{\mu 
\nu}^{ab},
	 \label{e43}
\end{align}
where 
\begin{align}
		\left({\mathfrak{D}_\text{\text{T}}}\right)_{\mu \nu}^{ab}  = & - 
\delta_{\mu \nu} \left({D^2}\right)^{ab} + 2i\bar{g} F_{\mu \nu}^{ab}, 
			\label{eq:DT2}
		\\
		\left({\mathfrak{D}_\text{L}}\right)_{\mu \nu}^{ab} = & - D_\mu^{ac} 
D_\nu^{cb} .
			\label{e42L}
\end{align}
The latter are further discussed in \cite{Reuter:1993kw, Reuter:1994zn, Reuter:1997gx}. 
Some useful properties of these kinetic operators for the present case are
\begin{align}
\left[{ \mathfrak{D}_\text{T},\mathfrak{D}_\text{L}}\right]_{\mu \nu}^{ab} & = 
0,
 \label{e44} \\
\left({\mathfrak{D}_\text{\text{L}}}\right)_{\mu \nu}^{ab} J_\nu^b &= 0,
\quad% \label{e45}\\
J_\mu^a  \left({\mathfrak{D}_\text{\text{L}}}\right)_{\mu \nu}^{ab} %&
= 0.
 \label{e46}
\end{align}
\Eqref{e44} relies on the assumption of covariantly constant background field. 
Equations \eqref{e46} are a direct consequence of the 
definition of the longitudinal kinetic operator given by \Eqref{e42L} 
 and of current conservation.

The source contribution to the Schwinger functional, \Eqref{e41}, now
takes the following form in the large-mass expansion
\begin{align}
W_{\text{source}}[A,v]  & = - \frac{1}{2} J_\mu^a \sum_{n=1}^{\infty} 
{\left[{\left({\frac{i}{\bar{m}}}\right)^{2n+2} 
\left({\mathfrak{D}_\text{T}^n}\right)_{\mu \nu}^{ab}}\right]} J_\nu^b 
\label{e48}
\end{align}
which highlights that the longitudinal contributions of the gluon propagator
and thus any gauge-parameter $\xi$ dependence drops out of the 1PR source term 
as a consequence of Eqs.~\eqref{e44} and \eqref{e46}. For the evaluation of 
\Eqref{e48}, we need the following commutator valid for covariantly constant 
background fields
\begin{align}
\left[{D^2, D_\mu}\right]^{ab}  = - 2i\bar{g} F_{\alpha \mu}^{ac} D_\alpha^{cb} 
. 
\label{e49}
\end{align}
Then, using the fact that the current has a longitudinal form, 
cf.~\Eqref{eq:longitudinalcurrent},  we obtain from the definition of $\mathfrak{D}_\text{T}$ in 
\Eqref{eq:DT2},
\begin{eqnarray}
 \mathfrak{D}_{\text{T}, \mu\nu}^{ab} J^b_\nu&=& -(D^2)^{ab} 
D_\mu^{bc}\chi^c +2i\bar{g}F^{ab}_{\mu\nu}J^b_\nu \nonumber\\
&=& D_\mu^{ab} D_\alpha^{bc} J_\alpha^c  - 2i\bar{g} F^{ab}_{\mu\nu}J^b_\nu
 + 2i\bar{g} F^{ab}_{\mu\nu}J^b_\nu \nonumber\\
 &=&0.
\end{eqnarray}
Here, the first term on the right-hand side vanishes due to
current conservation and the other two cancel.
Hence we conclude that the 1PR source contribution to the Schwinger functional
,
\begin{align}
\left. J_\mu^a \left({M^{-1}}\right)^{ab}_{\mu \nu} J_\nu^b\
\right|_{DF=0, DJ=0}  = 0 ,
\label{e50}
\end{align}
vanishes for covariantly constant backgrounds and for a conserved and longitudinal current 
$J[A,v]$ to all orders in the large-mass expansion. Note that the equations of 
motion for the background field are consistent with these assumptions, even 
entail them, but need not be satisfied in itself for the conclusion to hold.
Equation~\eqref{e50} also holds true  in the case where the 
current is given by \Eqref{eq:defJofv}. 
In fact in this case the current is
covariantly longitudinal, and in addition current conservation can be used
to show that $J=0$.
 
Alternatively, we could have imposed 
the consistency condition for the $v$ field, \Eqref{e25}, that arises from 
current conservation.
In fact, if this consistency condition is satisfied, then 
the auxiliary quantity $\chi^b$ introduced above vanishes identically and so 
does the current $J[A,v]$. In this case, \Eqref{e50} is satisfied as well 
independently of the background field, assuming the absence of relevant zero 
modes of the gluon propagator.

In either case, this leads to an expression for the one-loop Schwinger 
functional
\begin{equation}
\begin{aligned}
- W_{\text{1L}}[A]
\bigg|_{DF=0,D\cdot J=0} & =& S_\text{v}[A] - \ln\Delta_\text{FP}[A]  \\
&& + 
\frac{1}{2} \ln\det M[A], \label{e51}
\end{aligned}
\end{equation}
being independent of any $v$ contribution and thus of the nonlinear 
gauge-fixing sector except for the BRST invariant mass terms. As a consequence, 
one-loop $S$-matrix elements also remain independent of the $v$ field as 
expected.

%%%%%%%%%%%%%%%%%%%%%%%%%%%%%%%%%%%%%%%%%%%%%%%%%%%%%%%%%%%%%%%%%%%%%%%%%%
\section{Schwinger Functional with a disorder $v$ field }  \label{c4s2}
%%%%%%%%%%%%%%%%%%%%%%%%%%%%%%%%%%%%%%%%%%%%%%%%%%%%%%%%%%%%%%%%%%%%%%%%%%

The choice of a Fourier weight in the implementation of the gauge-fixing 
procedure, see Section \ref{s1s1}, leads to a $v$-dependent generating 
functional, \Eqref{e11}. The linear part of the gauge fixing condition, 
\Eqref{e16}, introduces a further source of $v$ dependence, which subsequently 
leads to 1PR terms in the Schwinger functional and also affects the 
Faddeev-Popov determinant. As shown before, one-loop results become independent 
again once consistency and/or on-shell conditions are used. However, from a 
practical viewpoint, it can be useful to have a computational formalism where no
implementation of consistency/on-shell conditions at certain stages is required.

It is useful to think of the $v$ field as a gauge-parameter field.
The most common implementation of the gauge-fixing condition is 
through a Gaussian averaging over the noise. The latter
would correspond to a Gaussian average over the $v$ field.
This averaging can either be implemented as an annealed or quenched disorder.
While the former prescription coincides with a standard Gaussian 
Nakanishi-Lautrup sector, the latter is less trivial and we therefore
find it interesting to analyze. In detail, we focus on the 
following quenched average over the external Nakanishi-Lautrup field 
of the Schwinger functional 
\begin{align}
W_{\text{1L}}[A] &= \mathcal{N} \int { \mathcal{D}v \; e^{- S_\NL[v]} \; W_{\text{1L}}[A,v]}, \label{e48b}\\
S_\NL[v]&= \int{\!
	\dd^d x\, \frac{v^2}{2 \alpha}},
\end{align}
where $\mathcal{N}$ is a normalization constant fixed by normalizing the 
free quenched average to unity.
 
Then, the terms that originate from the $v$ dependence and their 
effects can be studied. As only the last contribution in 
\Eqref{eq:Schwinger1L} is changed by this averaging,
it is useful to denote
\begin{equation}
	W_{\text{source}}[A]
	=-\frac{\mathcal{N}}{2}
	\int  \mathcal{D}v \; e^{- S_\NL[v]} \; W_{\text{source}}[A,v]. \label{qWsource}
\end{equation}
Note that a generic $v$ field as a representative of all configurations 
to be integrated over will generally not satisfy the consistency condition 
\eqref{e25} for a given background field. Whether or not such violations affect 
final results or may average out thus needs to be investigated. In the 
following, we will do so on the one-loop level where the $v$ 
field occurs only in the 1PR source term of the Schwinger functional, whereas 
the effective action remains unaffected.

With regard to the structure of the current $J[A,v]$ that carries the 
$v$ dependence inside the source term $W_\text{source}[A,v]$, cf.~\Eqref{e25}, 
we split the current into the derivative term and the term containing the ghost 
mass, and write the source term as,
\begin{widetext}
\begin{equation}
\begin{aligned}
W_\text{source}[A] = -  \left({\frac{1}{2}}\right)  \int  \mathcal{D}v \; \E^{- 
\int{\frac{v^2}{2 \alpha}}} & \left[{\left({D_\mu v}\right)^a 
\left({M^{-1}}\right)_{\mu \nu}^{ab} \left({D_\nu v}\right)^b}\right.  \qquad 
\qquad \qquad \qquad \qquad \qquad \; \; \; \; \; \; \text{(i)} \\ 
& \quad -\bar{m}_\text{gh}^2 \left({D_\mu v}\right)^a \left({M^{-1}}\right)_{\mu 
\nu}^{ab} \left({D_\nu \left({\frac{1}{D^2}}\right) v}\right)^b \qquad \qquad 
\qquad \; \text{(ii)} \\ 
& \quad  -\bar{m}_\text{gh}^2 \left({D_\mu \left({\frac{1}{D^2}}\right) 
v}\right)^a \left({M^{-1}}\right)_{\mu \nu}^{ab}  \left({D_\mu v}\right)^b 
\qquad \qquad \; \; \; \; \; \; \; \; \text{(iii)}\\ 
&  \quad \left.{+ \bar{m}_\text{gh}^4 \left({D_\mu \left({\frac{1}{D^2}}\right) 
v}\right)^a \left({M^{-1}}\right)_{\mu \nu}^{ab} \left({D_\nu 
\left({\frac{1}{D^2}}\right) v}\right)^b}\right]. \quad \; \; \;  \text{(iv)} 
\end{aligned}
\label{e63a}
\end{equation}
\end{widetext}
Performing the Gaussian functional integral, the four terms can be written as 
traces in coordinate, color and Lorentz space,
\begin{subequations}
\begin{align}
\text{(i)} &=  \left({\frac{\alpha}{2}}\right) \tr_{\text{xcL}}{\left[{D_\mu 
M^{-1}_{\mu \nu} D_\nu}\right]} \label{e64a}  
\end{align}
\begin{align}
\text{(ii, iii)} &= - \left({\frac{\alpha 
\bar{m}_\text{gh}^2}{2}}\right) \tr_{\text{xcL}}{\left[{\frac{1}{D^2}D_\mu 
M^{-1}_{\mu \nu} D_\nu}\right]} \label{e64b}
\end{align}
% \begin{align}
% \text{(iii)} &= - \left({\frac{\alpha \bar{m}_\text{gh}^2}{2}}\right) 
% \tr_{\text{xcL}}{\left[{\frac{1}{D^2}D_\mu M^{-1}_{\mu \nu} D_\nu}\right]} 
% \label{e64c}  
% \end{align}
\begin{align}
\text{(iv)} &=  \left({\frac{\alpha \bar{m}_\text{gh}^4}{2}}\right) 
\tr_{\text{xcL}}{\left[{\frac{1}{D^2} \frac{1}{D^2}D_\mu M^{-1}_{\mu \nu} 
D_\nu}\right]},  \label{e64d} 
\end{align}
\end{subequations}
where the cyclicity of the trace has been taken into account.  As the gluon 
propagator $M^{-1}$ is sandwiched between covariant derivatives, the 
transversal parts drop out and we obtain
\begin{subequations}
\begin{align}
\text{(i)} &=  \left({\frac{\alpha}{2}}\right) \tr_{\text{xcL}}{\left[{ 
\frac{1}{\bar{m}^2 - \left({\frac{1}{\xi}}\right) D^2 } D^2}\right]}  
\label{e65a} 
\end{align}
\begin{align}
\text{(ii, iii)} &= - \left({\frac{\alpha \bar{m}_\text{gh}^2}{2}}\right) 
\tr_{\text{xcL}}{\left[{ \frac{1}{D^2} \frac{1}{\bar{m}^2 - 
\left({\frac{1}{\xi}}\right) D^2 } D^2}\right]} \label{e65b}
\end{align}
% \begin{align}
% \text{(iii)} &= - \left({\frac{\alpha \bar{m}_\text{gh}^2}{2}}\right) 
% \tr_{\text{xcL}}{\left[{ \frac{1}{D^2} \frac{1}{\bar{m}^2 - 
% \left({\frac{1}{\xi}}\right) D^2 } D^2}\right]} \label{e65c} 
% \end{align}
\begin{align}
\text{(iv)} &=  \left({\frac{\alpha \bar{m}_\text{gh}^4}{2}}\right) 
\tr_{\text{xcL}}{\left[{ \frac{1}{D^2} \frac{1}{D^2} \frac{1}{\bar{m}^2 - 
\left({\frac{1}{\xi}}\right) D^2 } D^2}\right]}. \label{e65d} 
\end{align}
\end{subequations}
It is interesting to note that all terms are proportional to the width 
parameter $\alpha$ of the quenched disorder field; this implies that all terms 
vanish in the limit $\alpha\to 0$. In this limit, the amplitude of $v$ as a 
disorder field vanishes, such that the consistency condition \Eqref{e25} is 
evidently satisfied as also is current conservation, imposed explicitly in the 
preceding section. Still at this point, one may wonder whether the occurrence 
of 
inverse Laplacians may lead to ill-defined expressions.  

In order to obtain more explicit results, let us evaluate the traces for 
a covariantly constant, pseudoabelian magnetic field, as also used before, see Section \ref{sec:Gamma1L_beta}. 
It is already interesting to note for this case that none of the above 
traces are affected by the unstable Nielsen-Olesen mode which has dropped out as 
a consequence of the implicit longitudinal projection. For convenience, we 
introduce auxiliary functions that capture the dependence of the operators to be 
traced on the Laplacian. E.g., for the first term (i), we introduce
\begin{align}
h_{\text{(i)}} (x) =  - \frac{\xi x}{x+ \xi \bar{m}^2}, \label{e66a}
\end{align}
such that the trace in \Eqref{e65a} can be written as
\begin{align}
 \left.{\tr_{\text{xcL}} { \left[{h_{\text{(i)}} (x)}\right]}}\right|_{x=-D^2} = -
\left.{\tr_{\text{xcL}} {\left[{\frac{\xi x}{x+ \xi 
\bar{m}^2}}\right]}}\right|_{x=-D^2}. \label{e67a} 
\end{align}
The trace can be evaluated with heat-kernel techniques using the 
Laplace transform of the auxiliary function $h_{\text{(i)}}$ which leads us to 
the propertime representation. As before, we normalize the results by a constant 
subtraction such that the terms vanish at zero background field. For the trace, 
we find
\begin{equation}
\begin{aligned}
&\tr_{\text{xcL}} { \left[{h_{\text{(i)}} \left({-D^2}\right) - h_{\text{(i)}} 
\left({-\partial^2}\right)}\right]} =  - {\frac{\Omega_4 }{\left({4 
\pi}\right)^2}} \sum_{\ell=1}^{N^2-1} \\ 
&  \times { \int_0^{\infty}{ \frac{ds}{s^2} }} \left[{\left({\xi 
\bar{m}}\right)^2 \E^{-\xi \bar{m}^2 s} -  2 \xi  \delta(s)}\right] 
\left[{\frac{s B_\ell}{\sinh{s B_\ell}}-1}\right],
\end{aligned}
\label{e68a}
\end{equation}
where the $\delta$ function at the integral boundary is understood to 
contribute with half of its weight. We observe that the result is finite, the 
integral converges at both boundaries; moreover, it vanishes in the Landau 
gauge limit $\xi\to 0$. It is furthermore instructive to study the small-field 
limit at order $B^2$ which, on the level of the effective action, 
would correspond to 
the classical terms contributing to the renormalization of the charge and the 
field strength. It is straightforward to check that these possible lowest order 
terms of \Eqref{e68a} vanish, 
\begin{align}
\frac{1}{\Omega_4} \text{(i)}_{\text{vs}} =  0 + \mathcal{O}(B^4/\bar{m}^8), 
\label{e70a}
\end{align}
where the subscript 'vs' denotes vacuum subtraction.

The same computational steps can be applied to the terms (ii) and (iii), cf.~\Eqref{e65b}, using the auxiliary function 
\begin{align}
h_{\text{(ii)}}(x) = \frac{\xi}{x + \xi \bar{m}^2} .\label{e71a}
\end{align}
The corresponding vacuum subtracted trace yields
\begin{equation}
\begin{aligned}
& \tr_{\text{xcL}} { \left[{h_{\text{(ii)}} \left({-D^2}\right) - 
h_{\text{(ii)}} 
\left({-\partial^2}\right)}\right]} \\
&  =  \Omega_4 {\frac{ \xi}{\left({4 
\pi}\right)^2}} \sum_{\ell=1}^{N^2-1}  { \int_0^{\infty}{ \frac{ds}{s^2}  \E^{- 
\xi \bar{m}^2 s} \left[{\frac{s B_\ell}{\sinh{s B_\ell}}-1}\right]}}. 
\end{aligned}
\label{eq:hiitrace}
\end{equation}
Again, we observe that the integral expression is finite and vanishes 
in the Landau-gauge limit $\xi\to 0$. For this term, we obtain a contribution to quadratic order,
\begin{equation}
\begin{aligned}
& \frac{1}{\Omega_4} \text{(ii, iii)}_\text{vs} = \\
&  \quad \left({\frac{1}{4 
\pi}}\right)^2 \left({\frac{\bar{m}_\text{gh}}{\bar{m}}}\right)^2 
\left({\frac{\alpha}{12}}\right)  \Nc\, (\bar{g}B)^2  + 
\mathcal{O}(B^4/\bar{m}^8). \label{e72a}
\end{aligned}
\end{equation}
Note that the absence of the gauge parameter is a result  
of the expansion which is valid only for $\bar{g}B/(\xi\bar{m}^2)\ll 1$ and 
does not include the Landau-gauge limit.

For the fourth functional trace, \Eqref{e65d}, we introduce the auxiliary 
function
\begin{align}
h_{\text{(iv)}}(x) =  - \frac{\xi}{x \left({x+ \xi \bar{m}^2}\right)} \label{e73a}
\end{align}
which leads to the explicit form of the vacuum subtracted functional trace
\begin{equation}
\begin{aligned}
& \tr_{\text{xcL}} { \left[{h_{\text{(iv)}} \left({-D^2}\right) - 
h_{\text{(iv)}} 
\left({-\partial^2}\right)}\right]} = - {\frac{\Omega_4 }{\left({4 \pi 
\bar{m}}\right)^2}}\sum_{\ell=1}^{N^2-1} { |B_\ell|} \\
& \times{ \int_0^{\infty} \frac{dx}{x^2} \left[{1- \E^{- \frac{\xi 
\bar{m}^2}{|B_\ell|} x}}\right] \left[{\frac{x}{\sinh{x}}-1}\right]},
\end{aligned}
\label{e74a}
\end{equation}
where we substituted $x = s |B_\ell|$. Again, the integral is finite and 
vanishes in the Landau-gauge limit $\xi\to 0$. As an interesting feature, we 
observe that the integral in \Eqref{e74a} approaches a finite constant in the 
small-field limit. Up to quadratic order, the expansion yields
\begin{align}
& \frac{1}{\Omega_4} \text{(iv)}_\text{vs} = - \left({\frac{1}{4 \pi}}\right)^2 
\left({\frac{\bar{m}_\text{gh}}{\bar{m}}}\right)^4 \left({\frac{\alpha}{ 
2}}\right) \label{eq:ivexp}\\
&\times \left[ \bar{m}^2 \ln \left(\frac{1}{2}\right) \sum_{\ell=1}^{N^2-1} 
|B_\ell|   + \frac{1}{6\xi} \Nc (\bar{g}B)^2 \right] + \mathcal{O}(B^4/\bar{m}^8).  
\nonumber
\end{align}
The occurrence of the term linear in $B_\ell \sim \bar{g}B$ is noteworthy as 
this term would be invisible in a small-field expansion which is an 
expansion in $(\bar{g}B)^2$.
As before, the dependence of the 
quadratic term on $1/\xi$ is a result of the expansion which becomes invalid in 
the Landau-gauge limit. The full expression vanishes in the Landau gauge 
$\xi=0$. 

\par Collecting all results which were derived for the functional traces in the 
weak-field expansion, \Eqref{e70a}, 
\Eqref{e72a} \& \Eqref{eq:ivexp} and inserting into the source contribution to the Schwinger functional, 
\Eqref{e63a}, then
\begin{equation}
\begin{aligned}
& W_\text{source}[A]   \\
& = \frac{\Omega_4}{\left({4 \pi}\right)^2}
\left({\frac{\bar{m}_\text{gh}}{\bar{m}}}\right)^2 \left({\frac{ \alpha}{2} }\right) \left[{- \bar{m}^2 \ln{\left({\frac{1}{2}}\right) \sum_{l=1}^{N^2-1} \left|{B_\ell}\right|}}\right. \\
 & \qquad \qquad \qquad  \qquad \qquad  \quad + \left.{ \frac{\bar{g}^2 N_c}{6} \left({1- \frac{\bar{m}_\text{gh}^2}{\xi \bar{m}^2}}\right) B^2}\right] .
 \label{e78a}
\end{aligned}
\end{equation}
We observe that only the linear $B$-field term remains for 
the special choice of $\bar{m}_\text{gh}^2 = \xi \bar{m}^2$. 

The one-loop Schwinger functional takes the form
\begin{equation}
\begin{aligned}
- W_{\text{1L}}[A]  =&  S_\text{v}[A] - \ln\Delta_\text{FP}[A]  + 
\frac{1}{2} \ln\det M[A] \\
&- W_\text{source}[A].
\label{eq:Schw1loopque}
\end{aligned}
\end{equation}

In summary, we find that a treatment of the $v$ field as a quenched 
disorder field yields finite contributions to the 1PR source term in the 
Schwinger functional; of course, by averaging over $v$ this term becomes 1PI, 
as the quenched average corresponds to connecting the $v$-field legs.
The implicit violation of the consistency condition 
\eqref{e25} does not immediately cancel in the disorder average. Nevertheless 
we observe that these contributions can be controlled in various ways: first 
and importantly, they stay finite despite the nonlocal structure introduced by 
inverse Laplacians, second they vanish in the limit of both $\alpha\to 0$ 
(vanishing disorder amplitude) as well as $\xi\to 0$ (Landau gauge). All terms 
remain subdominant to the one-loop action or even vanish in the large-field 
limit and can also be 
controlled in a small-field 
expansion, even though
	the appearance of a finite
	term linear in $|B|$
signals the presence
of non-analyticities at
vanishing field strength. 

For the case of a constant magnetic 
field, also the Nielsen-Olesen unstable mode does not play any role. Provided 
that these properties translate to higher-loops or even nonperturbative 
computations, the treatment of the $v$ field as a 
disorder field can be a useful computational strategy to deal with this gauge 
degree of freedom.

%%%%%%%%%%%%%%%%%%%%%%%%%%%%%%%%%%%%%%%%%%%%%%%%%%%%%%%%%%%%%%%%%%%%%%%%%
\section{$v$-Field Contributions to the Connected Background 2-point 
Function} 
\label{s1s5}
%%%%%%%%%%%%%%%%%%%%%%%%%%%%%%%%%%%%%%%%%%%%%%%%%%%%%%%%%%%%%%%%%%%%%%%%%

Whereas the one-loop Schwinger functional is independent of the $v$ field on-shell, 
off-shell contributions to the connected background 2-point function are 
expected to appear from the 1PR source term, \Eqref{eq:Wsource}. In the 
following, we determine these contributions explicitly in the limit of 
vanishing background $A \rightarrow 0$ and external source 
$\jj\to 0$, see \Eqref{eq:defofj} for the definition of $j$. I.e., we 
study
\begin{equation}
 \Gv = \frac{\delta^2 W_{\text{source}}[j,A,v]}{\delta A\delta 
A}\Big|_{A,j\to0}, \label{eq:Gv}
\end{equation}
where $W_{\text{source}}[j=0,A,v]= \frac{1}{2}\mathcal{K} M^{-1}\mathcal{K}$, 
cf.~\Eqref{eq:WsAV}. Since $\mathcal{K}=DF+J$, the limit of vanishing 
background field yields for this quantity using \Eqref{eq:defJofv},
\begin{equation}
 \mathcal{K}_\mu^a\big|_{A\to0} = J_\mu^a\big|_{A\to 0}  = \partial_\mu \left({1 
+ \frac{\bar{m}_\text{gh}^2}{-\partial^2}}\right) v^a, \label{e59}
\end{equation}
which corresponds to the external current discussed in 
Ref.~\cite{Asnafi:2018pre}. Now, current conservation implies the consistency 
condition \eqref{e25}, requiring the $v$ field to satisfy the massive 
Klein-Gordon equation in absence of a background
\begin{align}
0 = \left.{D_\mu^{ab} J_\mu^{b}}\right|_{A \rightarrow 0}  = \left({\partial^2 - 
\bar{m}_\text{gh}^2}\right) v^a .\label{e60}
\end{align}
Combining Eqs.~\eqref{e59} and \eqref{e60} tells us that the source term 
vanishes in absence of a background
\begin{equation}
 \mathcal{K}_\mu^a\big|_{A\to0} = J_\mu^a\big|_{A\to 0}  = 0. \label{e59b}
\end{equation}
Correspondingly, the functional derivatives of \Eqref{eq:Gv} acting on the 1PR 
source term, \Eqref{eq:WsAV}, only yield finite contributions in the $A\to0$ 
limit as long as they act on the $\mathcal{K}$ factors. A crucial building 
block is given by $\delta J/\delta A$ which is derived in the Appendix in 
\Eqref{d11}. Furthermore, we need the gluon propagator $M^{-1}$ for vanishing 
fields which, in our gauge, reads schematically
\begin{equation}
M^{-1}\big|_{A \rightarrow 0} =
\left({\frac{1}{\bar{m}^2 - \partial^2}}\right) \Pi_{\text{T}} +
\left({\frac{1}{\bar{m}^2- \frac{\partial^2}{\xi}}}
\right) \Pi_{\text{L}},
\label{e58}
\end{equation}
where we have used the longitudinal and transversal projectors
\begin{equation}
\left({\Pi_\text{L}}\right)_{\mu \nu}^{ab} = \left({\frac{\partial_\mu 
\partial_\nu}{\partial^2}}\right) \delta^{ab} , \quad 
\left({\Pi_\text{T}}\right)_{\mu \nu}^{ab} = \delta_{\mu\nu}\delta^{ab} -
\left({\Pi_\text{L}}\right)_{\mu \nu}^{ab}. 
\label{e56}
\end{equation}
The building blocks for the 2-point function contributions from the source term 
using the preceding results in momentum space together with \Eqref{d11} thus 
read,
\begin{align}
\frac{1}{2}&\left.{\frac{\delta^2\left[{J_\mu^a 
\left({M^{-1}}\right)^{ab}_{\mu \nu} J_\nu^b}\right]}{\delta A_{\alpha}^c (p_1) 
\delta A_{\beta}^d (p_2)} }\right|_{A \rightarrow 0} \label{e65}\\
&=  \bar{g}^2 
f^{abc} f^{aed}  
 \int_q  \frac{\left({2q+p_1}\right)_\alpha}{\bar{m}^2 + \frac{q^2}{\xi}} 
\frac{\left({2q -p_2}\right)_\beta}{q^2}  
 v_{-q-p_1}^e v_{q-p_2}^b  ,
\nonumber\\
%\end{eqnarray}
%
%\begin{eqnarray}
\frac{1}{2}&
\left.{\frac{\delta^2\left[{\left({DF}\right)_\mu^a 
\left({M^{-1}}\right)^{ab}_{\mu \nu} \left({DF}\right)_\nu^b}\right]}{\delta 
A_\alpha^c(p_1) \delta A_\beta^d(p_2)} }\right|_{A \rightarrow 0} \label{e66}\\
&= \delta^{cd} 
\frac{p_1^2}{\bar{m}^2+p_{{1}}^2}
 \left({p_1^2 \delta_{\alpha \beta} - p_{1 \alpha} p_{1 \beta}}\right) 
\delta_{p_1, -p_2}, \nonumber \\
&\!\!\!\!\!\left.{\frac{\delta^2\left[{J_\mu^a \left({M^{-1}}\right)^{ab}_{\mu \nu} 
\left({DF}\right)_\nu^b}\right]}{\delta A_\alpha^c(p_1) \delta A_\beta^d(p_2)} 
}\right|_{A \rightarrow 0}  
%=\left.{\frac{\delta^2\left[{\left({DF}\right)_\mu^a 
%\left({M^{-1}}\right)^{ab}_{\mu \nu} J_\nu^b}\right]}{\delta A_\sigma^d(p_1) 
%\delta A_\rho^c(p_2)} }\right|_{A \rightarrow 0} 
= 0 .\label{e67}
\end{align}
Whereas the $v$-independent contribution in \Eqref{e66} is diagonal in 
momentum space, the $v$-dependent one from \Eqref{e65} is not. These nondiagonal 
parts parametrize the momentum-influx into the 2-point function provided by a 
possibly space-time dependent $v$ field. The sum of Eqs.~\eqref{e65} and 
\eqref{e66} represents the general result for the contributions from the 
1PR source term of the Schwinger functional to the 2-point correlator at this 
order for $v$ fields that satisfy the consistency condition, i.e. solve the 
massive Klein-Gordon equation.

Although we assumed current conservation, it is interesting to
	average \Eqref{e65} and \Eqref{e66} over $v$ with a Gaussian distribution, constrained by the requirement that the 
Klein-Gordon equation, \Eqref{e60}, holds. The latter constraint
can be implemented by means of a Lagrange multiplier field $\lambda^a$.
Hence we integrate both terms by means of the formulas 
\begin{align}
	& \braket{1} = \mathcal{N}\int\!\!\mathcal{D}\lambda^a\mathcal{D}v^a 
	\E^{\I\lambda^a (-\partial^2+\bar{m}_\gh^2)v^a-\frac{v^a v^a}{2\alpha}}
	\!=\! 1, \label{eq:integrvfield1}
	\\
	&\braket{v^a_p v^b_q} = \mathcal{N} \int\!\!\mathcal{D}\lambda^a\mathcal{D}v^a 
	\E^{\I\lambda^a (-\partial^2+\bar{m}_\gh^2)v^a-\frac{v^a v^a}{2\alpha}}
	v^a_p v^b_q= 0. \label{eq:integrvfield2}
\end{align}
where the normalization constant $\mathcal{N}$ is fixed by the requirement of~\Eqref{eq:integrvfield1}.

After averaging over the $v$ field, \Eqref{e65} will not contribute to the 
$v$-independent 2-point function, as a result of \Eqref{eq:integrvfield2}. 
The remaining term arising from \Eqref{e66} then yields  
the source contribution to the averaged 2-point function
\begin{align}
\braket{\Gv (p_1,p_2)}^{cd}_{\alpha\beta} = 
\frac{\delta^{cd} p_1^4}{\bar{m}^2+p_{{1}}^2}
 \left( \delta_{\alpha \beta} - \frac{p_{1 \alpha} p_{1 \beta}}{p_1^2}\right) 
\delta_{p_1, -p_2} \label{eq:vindepcorr}
\end{align}
The inverse $(\Gv)^{-1}$ of this function is a propagator type quantity. 
For this, we observe that $(\Gv)^{-1}\sim 1/p_1^{{2}}$ decays 
as usual in momentum space for large momenta 
implying that $S$-matrix 
contributions in the perturbative domain are not enhanced. 
At the same time, the $\sim 1/p_1^4$ 
behavior at small momenta is reminiscent to infrared slavery going hand in hand 
with the mass gap of the background field excitations.

Let us go back to the non-averaged case where, for concreteness, we
consider the special case of a homogeneous $v$ field
\begin{equation}
v_{q}^a  = v^a \sqrt{\left({2 \pi}\right)^d} \delta(q) .
\label{eq:homv}
\end{equation}
In this case, the consistency condition \eqref{e60} can be satisfied 
only for a vanishing ghost mass $\bar{m}_\text{gh} = 0$ which we assume here 
in addition. The final result for $\Gv$ then is diagonal in momentum space and 
reads
\begin{eqnarray}
\Gv&& (p|v)^{cd}_{\alpha \beta} 
  =   \bar{g}^2 f^{abc} f^{aed}  v^b v^e \delta_{p_1, -p_2}
\nonumber\\
&&\times  \left\{{\left[{\delta_{\alpha \beta} - {\frac{p_{1 \alpha} 
p_{1 \beta}}{p_1^2}}}\right] {\frac{1}{\bar{m}^2 +p_1^2}} }
%\right. \\  & \left.
{+ {\frac{p_{1\alpha} p_{1 \beta}}{p_1^2}} {\frac{1}{\bar{m}^2 + \frac{p_1^2}{\xi}}} 
}\right\} \nonumber\\
&&+\ \delta^{cd} \frac{p_1^4}{\bar{m}^2+p_1^2} \left({\delta_{\alpha 
\beta} - \frac{p_{1 \alpha} p_{1 \beta}}{p_1^2}}\right) \delta_{p_1, -p_2}.
\label{e69}
\end{eqnarray}
For general gauge parameter, we observe transversal and longitudinal 
contributions; pure transversality occurs in the Landau gauge where 
longitudinal parts decouple completely. The $v$-dependent parts affect only 
modes which are orthogonal to the $v$ field in color space. 

Equation~(\ref{e69}) can also be averaged 
	over all possible
directions of the constant vector $v^a$, with a 
Gaussian weight. For this kind of average,
which is more constrained than the one considered in 
Eqs.~(\ref{eq:integrvfield1}) and (\ref{eq:integrvfield2}),
also the first term
on the r.h.s.~of \Eqref{e69}
would bring a nonvanishing contribution.
Notice however that the latter would remain finite 
and would not modify the structure of
one-loop renormalization constants.
By contrast, in
Ref.~\cite{Asnafi:2018pre}, constant $v^a$ fields have been observed
to yield nonvanishing contributions to the running of
the wave function renormalization of the fluctuation field,
in absence of a background, with or without averaging over
$v^a$.

%%%%%%%%%%%%%%%%%%%%%%%%%%%%%%%%%%%%%%%%%%%%%%%%%%%%%%%%%%%%%%%%%%%%%%%%%
\section{$v[A]$-Field Contributions to the Connected Background 2-point 
Function} \label{s1s6}
%%%%%%%%%%%%%%%%%%%%%%%%%%%%%%%%%%%%%%%%%%%%%%%%%%%%%%%%%%%%%%%%%%%%%%%%%

The requirements of current conservation implied by the background equations of 
motion impose the consistency condition \eqref{e25} on the $v$ field. So far, 
we have implemented this condition at various stages of our studies. In 
particular in the preceding section, we have made use of the consistency 
condition, once we have varied the 1PR source term of the Schwinger functional 
with respect to the background field. Alternatively, we may assume that the 
consistency condition is satisfied from the beginning, rendering the $v$ field 
a functional of the background, $v=v[A]$. Correnspondingly, the connected 
background 2-point function will receive additional contributions, as is 
analyzed in the following. 

The difference to the result from the previous section arises from the 
derivative of the current,
% %
% \begin{equation}
%  \frac{\delta J[A,v]}{\delta A}= \frac{\delta J[A,v]}{\delta A}\Big|_{v} + N,
%  \label{e70}
% \end{equation}
% %
%
\begin{equation}
\frac{\delta J[A,v[A]]}{\delta A}= \frac{\delta J[A,v]}{\delta A}\Big|_{v=v[A]} + N,
	\label{e70}
\end{equation}
where the first term on the right-hand side first appeared in the 
preceding section and was computed in \eqref{d11}. 
Denoting
\begin{align}
	\mathrm{v}^a&=v[A\to0],\qquad 
	\frac{\delta  \mathrm{v}^a}{\delta A_{\alpha}^c}
    =\frac{\delta {v}^a}{\delta A_{\alpha}^c}\Bigg|_{A\to0},
\end{align}
the additional term $N$ for $A\to 0$
reads
\begin{eqnarray}
\left({N_{\mu \alpha}^{ac}}\right)_{xy}\Big|_{A\to0} &=& \partial_\mu^x 
\left({\frac{\delta  \mathrm{v}_x^a}{\delta A_{\alpha y}^c}}\right) 
\label{e71}\\
&&- \bar{m}_{\text{gh}}^2 \int_w 
{\left({\frac{1}{\partial^2}}\right)_{xw} \; \partial_\mu^w \left({\frac{\delta  
\mathrm{v}_w^a}{\delta A_{\alpha y}^c}}\right)}. \nonumber
\end{eqnarray}
This $N$ term gives rise to additional contributions to the connected 
background 2-point correlator not accounted for in the preceding section. The 
necessary functional derivative $\delta v/\delta A$ can be extracted from the 
consistency condition
\begin{equation}
( D^2 - \bar{m}_\text{gh}^2) v[A]=0.
\label{eq:conscondagain}
\end{equation}
Taking the functional derivative of this equation with respect to the 
background, yields in the limit $A\to 0$
\begin{align}
\int_z { \left.{\frac{\delta \left({D^2}\right)^{ab}_{xz}}{\delta A_{\alpha 
y}^c}}\right|_{A \rightarrow 0} \mathrm{v}_z^b}
= (\partial_x^2- \bar{m}_\text{gh}^2) 
\left({\frac{\delta 
\mathrm{v}_x^a}{\delta A_{\alpha y}^c}}\right). \label{e73}
\end{align}
Multiplying both sides by the kernel of the Klein-Gordon operator allows us to 
solve for the desired functional derivative, which in momentum space reads
\begin{equation}
\left.{\frac{\delta v^a_q}{\delta A_{\alpha p}^c}}\right|_{A \rightarrow 0} = - i 
\bar{g} f^{abc} \left({\frac{1}{\bar{m}_\text{gh}^2+q^2}}\right) 
\left({2q-p}\right)_\alpha \mathrm{v}_{q-p}^b.
\end{equation}
Upon insertion into \Eqref{e71} and subsequently into \Eqref{e70}, the 
essential building block for the 2-point function is the current derivative 
which now becomes
\begin{equation}
\left.{\frac{\delta J_{\mu q}^a}{\delta A_{\alpha p }^c}}\right|_{A \rightarrow 
0} \!\!\! = - \bar{g} f^{abc} {\frac{q_\mu 
\left({2q-p}\right)_\alpha}{q^2}}  %\\
%\times 
\left[{1- \frac{p^2-2p \cdot q}{\bar{m}_\text{gh}^2+q^2}}\right] 
\mathrm{v}_{q-p}^b. \label{eq:dJdAvA}
\end{equation}
This equation replaces \Eqref{d11} as used in the preceding section. 
Correspondingly, using \Eqref{eq:dJdAvA} in the analogue of Eqs.~\eqref{e65} and 
\eqref{e66}, we arrive at the following result for the background 2-point 
function arising from the 1PR source term
\begin{widetext}
\begin{eqnarray}
\Gv (p_1,p_2|v)^{cd}_{\alpha\beta} 
 &=& \delta^{cd} 
\frac{p_1^2}{\bar{m}^2+p_{1}^2}
 \left({p_1^2 \delta_{\alpha\beta} - p_{1 \alpha} p_{1 \beta}}\right) 
\delta_{p_1, -p_2}+ \bar{g}^2 f^{abc} f^{aed}  \int_q{ 
\left({\frac{1}{\bar{m}^2+ \frac{q^2}{\xi}}}\right) 
\left({\frac{1}{q^2}}\right) \left({2q+p_1}\right)_\alpha} 
\left({2q-p_2}\right)_\beta\nonumber \\
&& \!\times \left[ \!\left({\frac{1}{\bar{m}_\text{gh}^2+q^2}}\right)^2 
\left({p_1^2+2p_1 \cdot q}\right) \left({p_2^2-2 p_2 \cdot q}\right)
%\right. 
%\\
%&& \qquad 
- \frac{(p_2^2-2 p_2 \cdot q)}{\bar{m}_\text{gh}^2+q^2}  
%\\
%&& \left.{ \qquad
 + \frac{(p_1^2+2p_1 \cdot q)}{\bar{m}_\text{gh}^2+q^2}  + 1\right] 
\!\mathrm{v}_{q-p_2}^b \mathrm{v}_{-q-p_1}^e.\nonumber\\
\label{e76}
\end{eqnarray}
\end{widetext}
Again, these contributions parametrize the momentum influx that can be provided 
by a spacetime-dependent $v$ field. It is interesting to observe that the 
additional terms arising from treating the $v$ field as background dependent 
$v=v[A]$ go along with multiplicative ghost-propagator contributions, as a 
consequence of the consistency condition. 

\begin{table*}[t!]
\begin{center}
\begin{tabular}{ |c||c|c|c| } 
 \hline
 \multicolumn{4}{|c|}{\textsc{Different Approaches}} \\ 
 \hline 
 \shortstack{  Section}  &  \shortstack{  \ref{s1s4} }
  &   \ref{s1s5} &  \ref{s1s6}  \\
 \hline 
 \shortstack{ \parbox{2cm}{Background \\ (A)}}  &    \parbox{2cm}{$ D_\mu F_{\mu \nu} = 0 $ }    &    $  A \rightarrow 0$  &  $ A \rightarrow 0$ \\ 
 \hline 
 $ v$ - field & \parbox{3cm}{\begin{align*}
      \left({D^2-\bar{m}_\text{gh}^2 }\right) v = 0 
   \end{align*}} & \parbox{3cm}{\begin{align*}
     \left({\partial^2 - \bar{m}_\text{gh}^2}\right) v = 0 
   \end{align*}}  & \parbox{3cm}{\begin{align*}
     { \left({D^2 - \bar{m}_\text{gh}^2}\right) v[A]} = 0 
   \end{align*}} \\ 
    \hline
 $ {M^{-1}[A]}$ & \parbox{1cm}{\Eqref{e40}} 
   
   &  \Eqref{e58} 
   
   & \Eqref{e58} \\
   \hline
	$ W_\text{source}[A]$  
 	& 0 &  0 &  0 \\
 	\hline
 	 \parbox{2cm}{$  \Gv (p_1,p_2|v) $} & $-$
	 &  \parbox{3.5cm}{\Eqref{e65} $+$ \Eqref{e66}} &
   \Eqref{e76} \\
   \hline
	$ \braket{\Gv (p_1,p_2|v)}$  
 	& $-$ &  \Eqref{eq:vindepcorr} &  \Eqref{eq:vAindepcorr} \\
 	\hline
 	 \parbox{2cm}{$  \Gv  (p|v) $} & $-$
	 &  \Eqref{e69} &
   \Eqref{eq:GvvA} \\
   \hline
\end{tabular}
\caption{Summary of the approaches adopted in the evaluation of the new source-like contributions to the background connected correlation
	functions, associated to the special gauge-fixing sector constructed in Sec.~\ref{s1s1}.
	These contributions depend on two external
	fields: the background gluon field $A$, and
	the Nakanishi-Lautrup field $v$, which have been chosen
	as specified in the second and third rows respectively. 
	Notice that the $v[A]$ of the third column
	differs from the $v$ of the second column, in
	that we consider the $v$ field as a functional 
	of $A$ {\it before} taking functional
	derivatives w.r.t.~an arbitrary $A$. 
The main actor in the evaluation of the source-like contributions is the gluon propagator $M^{-1}$, for
which we refer to the corresponding explicit expression.
Under each assumption considered, we report the contributions to the zero-point function $ W_\text{source}[A]$ and to the
two-point function $\Gv (p_1,p_2|v)$.
Special forms of the latter have been obtained, either by averaging over the $v$ field with a constrained Gaussian
distribution, or by assuming $v$ to be constant.
The corresponding results are respectively recalled in the last two rows. 
The empty entries in the lower-left corner
correspond to computations of the two-point correlator in 
non-vanishing backgrounds and are left for future studies.} 
\label{t1}
\end{center}
\end{table*}

Similar to the previous section, let us now average over the $v$ field 
in the 2-point function, \Eqref{e76}, under the constraint of current conservation. 
Then, the average of the second term of \Eqref{e76} vanishes due to 
\Eqref{eq:integrvfield2} and the 
first term will contribute the same result as in \Eqref{eq:vindepcorr}, i.e.
\begin{align}
\braket{\Gv (p_1,p_2|v)}^{cd}_{\alpha\beta} = \delta^{cd} 
\frac{p_1^2}{\bar{m}^2+p_{{1}}^2}
 \left({p_1^2 \delta_{\alpha \beta} - p_{1 \alpha} p_{1 \beta}}\right) 
\delta_{p_1, -p_2}. \label{eq:vAindepcorr}
\end{align}
Therefore, the average over the $v$ field on the level of the 2-point 
function yields the same result for the 2-point function regardless of the 
initial consideration of a background-dependent or background-independent $v$ 
field.

For a simplified direct comparison with the results of the preceding section, 
it is again instructive to consider a homogeneous $v$ field, cf.~\Eqref{eq:homv}, leading us to the result
\begin{eqnarray}
\Gv&& (p|v)^{cd}_{\alpha \beta} 
  = \bar{g}^2 f^{abc} f^{aed}  v^b v^e \delta_{p_1, -p_2}
\nonumber\\
&&\times  \left\{{\left[{\delta_{\alpha \beta} - {\frac{p_{1 \alpha}
p_{1 \beta}}{p_1^2}}}\right] {\frac{1}{\bar{m}^2 +p_1^2}} }
%\right. \\  & \left.
{+ {\frac{p_{1 \alpha} p_{1 \beta}}{p_1^2}} {\frac{2}{\bar{m}^2 + \frac{p_1^2}{\xi}}} 
}\right\} \nonumber\\
&&+\  \delta^{cd} \frac{p_1^4}{\bar{m}^2+p_1^2} \left({\delta_{\alpha 
\beta} - \frac{p_ {1 \alpha} p_{1 \beta}}{p_1^2}}\right) \delta_{p_1, -p_2}.
\label{eq:GvvA}
\end{eqnarray}
Interestingly, the only change in comparison to \Eqref{e69} 
occurs in the longitudinal part in the form of a factor of 2. In the Landau 
gauge $\xi\to 0$, where this part decouples, the difference is irrelevant. 

In Table \ref{t1} we have summarized the main results found in 
Sections \ref{s1s4}, \ref{s1s5} \& \ref{s1s6}.

%%%%%%%%%%%%%%%%%%%%%%%%%%%%%%%%%%%%%%%%%%%%%%%%
\section{Summary}
\label{sec:conc}
%%%%%%%%%%%%%%%%%%%%%%%%%%%%%%%%%%%%%%%%%%%%%%%%

External-field methods and BRST-invariant perturbation
theory are two cornerstones of quantum field theory
and its applications to high energy theory.
Extending and generalizing these tools 
to aim at a 
description of phenomena such as 
color confinement or
the spontaneous breaking of symmetries,
most notably of chiral symmetry and scale invariance, 
and eventually of the physical particle spectrum,
requires new theoretical developments.

One successful semi-analytical method able to
bridge between the language of perturbation theory
and nonperturbative aspects of quantum and statistical
field theories
is Wilson's RG analysis of effective field
theories (EFTs). 
The power of the latter method is further enhanced by
the possibility to work in massive RG schemes, i.e.~to
assign a mass scale to each degree of freedom and to
use such scales as RG times, along whose flow 
the properties of the system continuously change
according to the EFT ideas of matching and decoupling.
Maybe the most classic embodiment and application of these 
ideas is the so-called functional RG (FRG),
in which the Wilsonian idea of a continuous RG flow of
effective theories is formulated at the level of an
exact functional equation for a mass-scale-dependent effective 
action~\cite{Wegner:1972ih,Polchinski:1983gv,Wetterich:1992yh,Morris:1993qb,Ellwanger:1993mw}.

The problem of introducing a Wilsonian floating mass scale in a classically 
scale-invariant gauge theory without breaking BRST symmetry is as old as 
relevant for contemporary applications.
In the FRG, it is re-phrased as the problem of constructing 
a BRST-preserving exact RG equation. While the latter problem has been 
tackled by several approaches in the literature, it is fair to say that none of 
them serves all purposes.
For these reasons, some of us re-considered this problem from a novel
perspective in  Ref.~\cite{Asnafi:2018pre}.
There, a BRST-preserving embedding of mass parameters for the 
fields of Faddeev-Popov-quantized Yang-Mills theory,
in absence of a background, was
achieved at the price of introducing an external 
Nakanishi-Lautrup field $v$, and some explicit nonlocality
in the ghost action.

The important question whether these unusual features
would compromise the consolidated perturbative
understanding of gauge-fixed Yang-Mills theory,
or might even radically change its one-loop
RG flow, was not completely answered there.
In this work, we have collected evidence at one loop that 
the novel gauge-fixing prescription 
does not impede a standard interpretation of the 
perturbative series, as well as a fruitful
implementation of the background field method.
In fact, after a straightforward generalization
of the gauge-fixing construction of Ref.~\cite{Asnafi:2018pre} to the special class of background-covariant gauges in
Secs.~\ref{s1s1} and \ref{s1s2},
we have shown that the one-loop structure of
the background Schwinger functional differs from the
standard one (corresponding to a Gaussian
Nakanishi-Lautrup action) only through a
one-particle-reducible source-like term
$W_\text{source}$ presented in Eqs.~\eqref{eq:Wsource}
and \eqref{eq:extcurr},
associated to the $v$-induced external 
color source $J$ of \Eqref{eq:defJofv}.

Thus, for an external Nakanishi-Lautrup field the
one-loop 1PI effective action
still comprehends just the standard
contribution of gluon and ghost loops.
Of course these loops, while giving rise to
the universal one-loop beta function
for the marginal gauge coupling (for the
derivation of which it was sufficient to
adopt dimensional regularization and the 
\MSbar scheme),
 are now featuring
arbitrary gluon and ghost mass parameters,
which affect both the one-loop divergences
and the background dependence of
the effective action, precisely as one would
expect for genuine IR-regulating mass
thresholds.
More precisely, in Sec.~\ref{sec:Gamma1L_beta}
we have deduced the threshold-depending
one-loop beta function in a propertime
regularization scheme, \Eqref{eq:betaPTmass}, and we have
illustrated how the presence of 
large-enough mass parameters
can cure the Nielsen-Olesen instability
at strong fields (see Fig.~\ref{fig:dimlessLagr})
for covariantly constant pseudo-Abelian
magnetic backgrounds.

This instability does not occur for a selfdual background. Within our 
approach, it turns out to be straightforward to include nonperturbative 
information about the gluon and ghost propagators in the Landau gauge in the 
form of the so-called decoupling solution. Using this input, we have found 
evidence that the effective action supports a gluon condensate beyond a 
critical coupling. This result serves as a first 
illustration that our approach can give immediate access to phenomenologically 
relevant quantities.

These computations serve as
examples of the use
and interpretation of the
BRST-invariant mass parameters.
As the latter enter through
the gauge-fixing sector, 
and therefore belong to BRST-exact
deformations of the classical
action, one might expect that
the associated scale symmetry breaking
remains confined in the
BRST-exact unphysical sector
of the theory space.
On the other hand,
quantum corrections lead to a dynamical
breaking of scale symmetry,
which is visible in
Figs.~\ref{fig:dimlessLagr}~and~\ref{fig:dualbackgroundcondensate},
in the form of special non-vanishing stationarity values of the field amplitude.
While in more traditional gauge-fixing schemes these
nontrivial saddle-points necessarily come along with the floating
regularization scale, as the bare action is scale-free,
in the present framework the latter scale is replaced by
the BRST-invariant mass parameters.
Thus, as a consequence of quantum corrections,
the tree-level breaking of scale invariance in the BRST-exact sector
acts as a seed which is communicated by radiative corrections
to the physical sector of the theory space.
Even more interestingly, we observe that this mechanism does not
require nonperturbative approximations or the discussion of
the Singer-Gribov ambiguity.

We have then devoted the rest of this study to the
investigation of the nontrivial contribution
$W_\text{source}$, as a functional of
both the background gluon field $A$ and
the external Nakanishi-Lautrup/disorder field $v$.
As far as the latter is concerned, it is reasonable
to assume color current conservation, such that 
the consistency condition
\Eqref{e25} holds true. 
This equation entails  a
mutual relation between the  two field configurations, which
can be either assumed to hold for arbitrary $A$ (off shell)
or for the chosen $A$ only (on shell).
The two assumptions correspondingly lead
to different structures in the derivatives of
$W_\text{source}$ w.r.t.~$A$,
depending on whether $v$ is 
a functional of $A$ or is
independent of it.
The results of our investigation are summarized in 
Table~\ref{t1}.

We have also further explored the idea of treating
$v$ as a disorder field, to be averaged over, which was
discussed also in Ref.~\cite{Asnafi:2018pre}.
The two alternative treatments of
annealed or quenched disorder then correspond to integrating out
the Nakanishi-Lautrup auxiliary field either first (at the level of the bare action) or last (at the level of the Schwinger functional).
Our findings for $W_\text{source}$ further substantiate
the general conclusion that at one-loop no major interpretational novelties appear, since the 
source contribution to the zero-point 
function is found to vanish, and the  
source contribution to the background
two-point function is finite
both for constant (i.e.~homogeneous) $v$ fields
and for Gaussian quenched disorder.

In the present study, we have taken advantage
of the mass parameters  which can be
included in the bare action of Yang-Mills theory
by means of the gauge-fixing sector
to describe the structure of one-loop corrections
to the pure background effective action, thus neglecting
nonvanishing sources (i.e.~expectation values)
for the gluon fluctuations and for the ghosts.
As a consequence, while we addressed the
running of the gauge coupling, we could not,
for instance, compute the running of the
mass parameters themselves.
Retaining nonvanishing sources besides the background field
would also allow for a discussion of the so-called
split Ward identities, namely of the symmetry 
corresponding to simultaneous shifts of the gluon background and of the
corresponding fluctuation.
This is especially relevant
for the construction of
functional truncations in the
FRG framework.
The latter is a possible interesting extension
of this work.

%\clearpage
%\thispagestyle{empty}

%%%%%%%%%%%%%%%%%%%%%%%%%%%%%%%%%%%%%%%%
\section*{Acknowledgments}
%%%%%%%%%%%%%%%%%%%%%%%%%%%%%%%%%%%%%%%%

We thank Shimasadat Asnafi and Kevin Falls for useful discussions.
This work has been funded by the Deutsche Forschungsgemeinschaft (DFG) under 
Grant Nos. 398579334 (Gi328/9-1) and 406116891 within the Research Training 
Group RTG 2522/1.
This project has also received funding from the European Union’s Horizon 2020 research and
innovation programme under the Marie Skłodowska-Curie grant agreement No 754496.

%%%%%%%%%%%%%%%%%%%%%%%%%%%%%%%%%%%%%%%%
\appendix
\section{Useful Relations for the connected background two-point function} 
\label{appd}
%%%%%%%%%%%%%%%%%%%%%%%%%%%%%%%%%%%%%%%%%%%%%%%%%%%%%%%%%%%%%%%%%%%%%%%%%

In this Appendix, we will summarize our conventions and some details needed 
for the computations in Sects.~\ref{s1s5} and \ref{s1s6}, where the 1PR source 
term contributions to the connected two-point function are studied. In the 
main text, we often use condensed notation in which color indices 
implicitly represent spacetime indices as well. 

For the transition from coordinate to momentum space, we use the Fourier 
conventions
\begin{equation}
\begin{aligned}
\Phi_{x}^i = \int_p e^{-ipx} & \Phi_{p}^i,  \quad   \Phi_p^i = \int_x {e^{ipx} \Phi_x^i}, \\
& \delta_{xy} = \int_p e^{-ip(x-y)}, 
\end{aligned}
 \label{d1}
\end{equation}
where 
\begin{equation}
\begin{aligned}
\int_p = \int{\frac{d^dp}{\left({2 \pi}\right)^d}}, \quad \int_x = \int{d^dx}.
\end{aligned}
\notag
\end{equation}
Using this more explicit but compact notation, the covariant derivative 
reads
\begin{align}
\left({D	_\mu^{ab}}\right)_{xy} = \left({\delta^{ab} \partial_\mu^x  + \bar{g} 
f^{acb} A_{\mu x}^c}\right)  \delta_{xy} \notag
\end{align}
reflecting a notation that is also used in the main text.

A building block required during the computations are functional derivatives of 
the Laplacian, which in position space reads
\begin{equation}
\begin{aligned}
\left({D^2}\right)^{ab}_{xy} =& \left[{\delta^{ab} \partial_x^2 + 2 \bar{g} 
f^{acb} A_{\mu x}^c \partial_\mu^x + \bar{g} f^{acb} \left({\partial_\mu^x 
A_{\mu x}^c}\right)}\right. \\ 
 & \; \left.{ + \bar{g}^2 f^{acd} f^{cbe} A_{\mu x}^d A_{\mu x}^e}\right] 
\delta_{xy} . 
\end{aligned}
 \label{d2}
\end{equation}
Specifically, we need its functional derivative for vanishing background fields,
\begin{align}
\left.{\frac{\delta \left({D^2}\right)^{ab}_{xy}}{\delta A^c_{ \alpha 
z}}}\right|_{A \rightarrow 0} = 2 \bar{g} f^{acb} \delta_{xz} 
\left({\partial_\alpha^x \delta_{xy}}\right) + \bar{g} f^{acb} 
\left({\partial_\alpha^x \delta_{xz}}\right) \delta_{xy}.  \label{d3} 
\end{align}
Also, the functional derivative of the inverse Laplacian operator at 
vanishing background is required. For this, we make use of the following 
relation for the functional derivative for the inverse of a generic operator,
\begin{align}
\frac{\delta \left({\Theta^{-1}}\right)^{ab}_{\mu \nu}}{\delta A_\alpha^c} = 
- \left({\Theta^{-1}}\right)^{ad}_{\mu \lambda} \left({\frac{\delta 
\Theta^{de}_{\lambda \rho}}{\delta A_\alpha^c}}\right)  
\left({\Theta^{-1}}\right)^{eb}_{\rho \nu}.  \label{d5} 
\end{align}
Using that the Laplacian at vanishing background is 
diagonal in momentum space, the derivative of the inverse 
Laplacian operator in momentum space at vanishing background reads
\begin{align}
\left.{\frac{\delta}{\delta A_{\alpha p}^c} \left({\frac{1}{D^2}}\right)^{ab}_{q_1 q_2}}\right|_{A \rightarrow 0}
= - i \bar{g}f^{acb} \frac{\left({q_1 + q_2}\right)_\alpha}{q_1^2 q_2^2} 
\delta_{q_1q_2p},  
\end{align}
where we have abbreviated $\delta(q_1-q_2-p_1) = \delta_{q_1q_2p_1}$.

Next, we turn to the source term induced by the presence of the $v$ 
field. In the main text, we discussed that the conservation of this source 
term, \Eqref{e25}, imposes a consistency condition on the $v$ field, which --  
for vanishing backgrounds -- corresponds to a  
massive Klein-Gordon equation, 
\begin{align}
\left.{\left({{D}_\mu^{ab}}\right)_{xy} J^b_\mu}\right|_{A \rightarrow 0} =0 
\Rightarrow \left({\partial^2_x - \bar{m}_\text{gh}^2}\right) v_x^a = 0, 
 \label{d7} 
\end{align}
which in momentum space takes the form 
\begin{align}
\left({q^2 + \bar{m}_\text{gh}^2}\right) v^a_q = 0.  \label{d8}
\end{align}
Correspondingly, the limit of vanishing backgrounds also allows us to write 
down explicit expressions for the source term in momentum space:
\begin{align}
\left.{J_{\mu q}^a}\right|_{A \rightarrow 0} = 
	\begin{cases}
	-i q_\mu \left({1+\frac{\bar{m}_\text{gh}^2}{q^2}}\right) v^a(q), \; & \left.{D \cdot J}\right|_{A \rightarrow 0}  \neq 0 \\
	0, & \left.{D \cdot J}\right|_{A \rightarrow 0} = 0
	\end{cases}
	 \label{d9}
\end{align}
Here we have distinguished between the cases where the consistency condition 
\eqref{d8} is satisfied (implying current conservation) or not. In the same 
fashion, we can give explicit momentum space expressions for the 
functional derivative of the current as needed in the main text. If \Eqref{d8} 
is not imposed as a constraint, then
\begin{equation}
\begin{aligned}
\left.{\frac{\delta J_{\mu q}^a}{\delta A_{\alpha p} ^c}}\right|_{A \rightarrow 
0} =	-\bar{g} f^{abc} & \left[{\delta_{\mu \alpha} + \bar{m}_\text{gh}^2 
\frac{\delta_{\mu \alpha}}{\left({q-p}\right)^2}}\right. \\ 
 & \quad \left.{+ \bar{m}_\text{gh}^2 \frac{q_{\mu} 
\left({2q-p}\right)_\alpha}{q^2 \left({q-p}\right)^2}}\right] v^b_{q-p} , 
\end{aligned}
 \label{d10}
\end{equation}
whereas imposing the consistency condition \eqref{d8} leads to
\begin{align}
\left.{\frac{\delta J_{\mu q}^a}{\delta A_{\alpha p} ^c}}\right|_{A \rightarrow 
0} =  \bar{g} f^{abc}  q_\mu \left({2q-p}\right)_\alpha 
\left({\frac{1}{q^2}}\right) v^b_{q-p}.  \label{d11} 
\end{align}
The latter corresponds to the relation that is used in Sect.~\ref{s1s5}.

% For the inverse Laplacian at vanishing background, 
% \begin{align}
% \left.{\left({\frac{1}{D^2}}\right)^{ab}_{xy}}\right|_{A \rightarrow 0} & = 
% \delta^{ab} K(x-y), \tag{D.4} \label{d4}
% \end{align}
% where $K(x-y)$ obeys $\partial_x^2 K(x-y) = \delta_{xy} $ and corresponds to the 
% kernel of the massless propagator in position space. 

\bibliography{bibliography} 
\bibliographystyle{dimtest}

\end{document}